\documentclass[journal]{IEEEtran}

\usepackage{enumerate}
\usepackage{booktabs}
\usepackage{xcolor}
\usepackage{colortbl}
\usepackage{amssymb}
\usepackage{algorithmic}
\usepackage{amsmath} 
\usepackage{amsfonts}
\newtheorem{theorem}{Theorem}

\newtheorem{proposition}{Proposition}

\newtheorem{definition}{Definition}

\ifCLASSINFOpdf
\usepackage[pdftex]{graphicx}
\else
\usepackage[dvips]{graphicx}
\fi
\usepackage{epstopdf}
\usepackage{array}
\ifCLASSOPTIONcompsoc
\usepackage[caption=false,font=normalsize,labelfont=sf,textfont=sf]{subfig}
\else
\usepackage[caption=false,font=footnotesize]{subfig}
\fi
\usepackage{float} 
\usepackage{fixltx2e}
\usepackage{color}
\usepackage{xcolor}
\usepackage{textcomp}
\usepackage[normalem]{ulem}
\usepackage{cite}
\usepackage{multibib} 
\usepackage{nomencl}
\makenomenclature
\usepackage{etoolbox}
\renewcommand\nomgroup[1]{%
	\item[\bfseries
	\ifstrequal{#1}{P}{Physical Variables}{%
		\ifstrequal{#1}{M}{Mathematical Symbols}{%
			\ifstrequal{#1}{S}{Subscripts and Superscripts}{}}} ]}

\usepackage{hyperref}
\hypersetup{
	colorlinks=true,
	linkcolor=blue,
	filecolor=magenta,      
	urlcolor=cyan,}
\IEEEoverridecommandlockouts

\begin{document}
\bstctlcite{BSTcontrol}
\title{Set Theory-Based Safety Supervisory Control for Wind Turbines to Ensure Adequate Frequency Response}

\author{Yichen~Zhang,~\IEEEmembership{Student Member,~IEEE,}
	   M.~Ehsan~Raoufat,~\IEEEmembership{Student Member,~IEEE,}
        Kevin~Tomsovic,~\IEEEmembership{Fellow,~IEEE,}
        and~Seddik~M.~Djouadi,~\IEEEmembership{Member,~IEEE}
        \thanks{This work was supported in part by the National Science Foundation under the Grant ECCS-1711432, in part by CY 18 Science Alliance Joint Directed Research Development (JDRD) Award, and in part by the Engineering Research Center Program of the National Science Foundation and the Department of Energy under NSF Award Number EEC-1041877.
        
        The authors are with the Min H. Kao Department of Electrical Engineering and Computer Science, The University of Tennessee, Knoxville, TN 37996 USA (e-mail: yzhan124@utk.edu, mraoufat@utk.edu, tomsovic@utk.edu, mdjouadi@utk.edu).}}

\markboth{A\MakeLowercase{ccepted by} IEEE TRANSACTIONS ON POWER SYSTEMS \MakeLowercase{on} A\MakeLowercase{ugust}, 2018 (DOI: 10.1109/TPWRS.2018.2867825)}%
{Shell \MakeLowercase{\textit{et al.}}: Bare Demo of IEEEtran.cls for IEEE Journals}
\maketitle

\begin{abstract}
Inadequate frequency response can arise due to a high penetration of wind turbine generators (WTGs) and requires a frequency support function to be integrated in the WTG. The appropriate design for these controllers to ensure adequate response has not been investigated thoroughly. In this paper, a safety supervisory control (SSC) is proposed to synthesize the supportive modes in WTGs to guarantee performance. The concept, region of safety (ROS), is stated for safe switching synthesis. An optimization formula is proposed to calculate the largest ROS. By assuming a polynomial structure, the problem can be solved by a sum of squares program. A feasible result will generate a polynomial, the zero sublevel set of which represents the ROS and is employed as the safety supervisor. A decentralized communication architecture is proposed for small-scale systems. Moreover, a scheduling loop is suggested so that the supervisor updates its boundary with respect to the renewable penetration level to be robust with respect to variations in system inertia. The proposed controller is first verified on a single-machine three-phase nonlinear microgrid, and then implemented on the IEEE 39-bus system. Both results indicate that the proposed framework and control configuration can guarantee adequate response without excessive conservativeness. 
\end{abstract}
\IEEEpeerreviewmaketitle

\section{Introduction}
Due in part to the increasing penetration of converter-interfaced sources, such as, the wind turbine generator (WTG), total system inertia has been decreasing. The result can be inadequate system frequency response as a small power disturbance may lead to a large frequency excursion during the transient period, that is, the period of inertial and primary responses \cite{GE_SFR, NREL_EI_SFR, Lawrence_SFR, NREL_J}. This poor transient response can trigger unnecessary over or under-frequency relay actions even though the system has adequate capacity to reach a viable steady state \cite{freq_limit_storage}. Thus, maintaining the system frequency within the continuous operation zone, or so-called safety\footnote{The term \emph{safety} is adopted from the control literature and in this context means a well-defined and allowable operating region. A safe response means the trajectories of all concerned states stays within the defined safe limits.} limits, under a certain set of disturbances has become increasingly important \cite{freq_limit_storage, freq_limit_load, rate_MIT} and necessary to real-world power system operations to avoid unnecessary loss of generation and load \cite{quebec}. 

\subsection{Literature Review}
Numerous frequency supportive functions for WTGs have been studied, which can be divided into two categories. The most common and representative method is to provide an additional signal associated with the measured grid frequency deviation \cite{DeAlmeida2007,Gautam2011a,Arani2013}, or its differential \cite{Lalor2005a,Kayikci2009,Wu2013}, a mix of both \cite{Morren2006,Mauricio2009,Vidyanandan2013,Junyent-Ferre2015,VanDeVyver2016,Wilches-Bernal2016} , or a pre-specified reference signal \cite{Keung2009,Miao2014,Wang2018Novel,zyc_PEDG2018} (referred to as the power surge control) to the torque/power or speed reference value to be tracked \cite{Mauricio2009}. These methods can be referred to as the supplementary-signal based methods. The emulated primary response is associated with the frequency deviation, while the emulated inertial response is associated with the rate of change of frequency (RoCoF) \cite{zyc_ENERGYCON2018}, which can be generated by filtering the frequency through a washout filter \cite{Lalor2005a}. 

The other type of approaches is to mimic the power-angle relation of traditional synchronous generators by means of modifying either the phase-lock loop (PLL) \cite{he2017inertia,Hu2017} or the active power controller \cite{Zhang2010,Zhang2016e,zyc_MRC_TPS_2018}. The angle used by the Park's transformation for synchronization is no longer obtained through the vector alignment, but calculated using the swing dynamics. Thus, inertia, load-damping effect and droop characteristics can be provided \cite{Zhang2016e}.

\begin{figure}[t]
	\centering
	\includegraphics[scale=0.35]{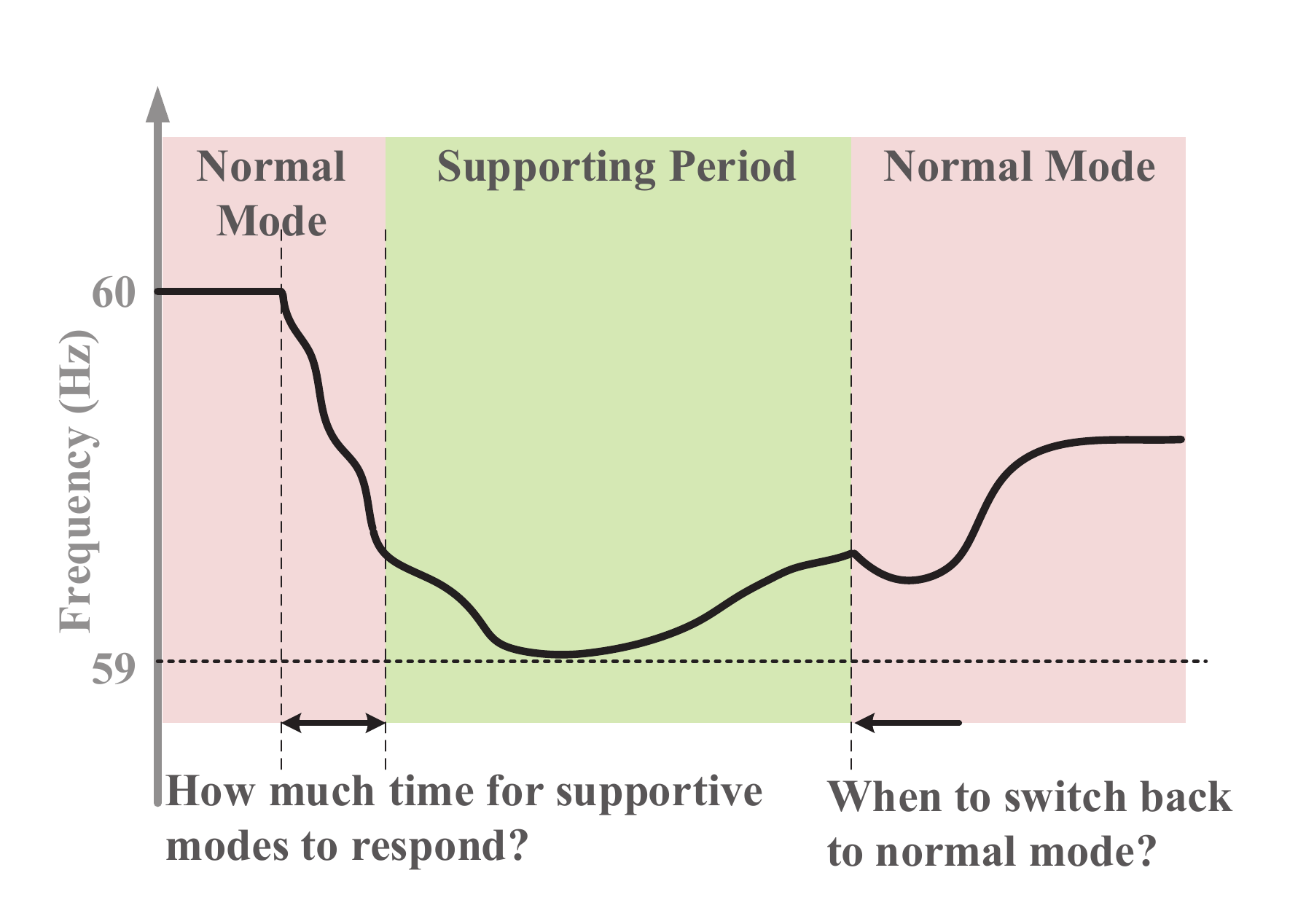}
	\caption{Challenges of synthesizing supportive modes in CIS as the switching instants between modes to achieve an adequate frequency response are still unclear.}
	\label{fig_Motivation_SSC}
\end{figure}

With all listed approaches, however, few research studies utilizing these functions to achieve adequate frequency response, i.e., bounded within the defined safety limits for a given set of contingency events. There are mainly two challenges that hamper good participatory research. The major challenge is the presence of deadbands in the supportive functions. The deadbands ensure that WTGs do not respond to small frequency fluctuations in the grid so as to extract maximum power from wind \cite{fundamental2016}. Since the power extraction is the primary tasks for WTGs, the deadbands in the supplementary loops are large compared to the traditional deadbands in governors. The WTGs integrated with supportive functions become hybrid dynamical systems, where the switching actions between modes to achieve an adequate response are not well understood to both industry and academia (as illustrated in Fig. \ref{fig_Motivation_SSC}  \cite{rate_MIT}). In addition, a fixed deadband may not be able to handle a high renewable penetration condition. Due to the stochastic and intermittent nature of renewable resources, the commitment of traditional plants will need to change dramatically over time, which could significantly change the system frequency response characteristics \cite{Teng2016a}\cite{Prakash2017}. For example, during the year 2012, several occasions took place in Germany, where around 50\% of overall load demand was covered by wind and PV units for a few hours. The regional inertia within the German power system changes dramatically between lower and higher levels \cite{Ulbig2014}. The second challenge for the existing controllers in WTGs is a lack of situational awareness capability as only terminal measurements are used. 
\par

Ref. \cite{freq_limit_storage} proposed to use energy storage systems to avoid unnecessary under-frequency load shedding (UFLS). A composite model of system frequency response (SFR) and energy storage systems is built to evaluate the frequency response with support. The switching threshold is determined by the initial RoCoF after the disturbance, which is very difficult to calculate. A commitment strategy for interruptible load to ensure adequate response is proposed in \cite{freq_limit_load}. The frequency nadir information under different commitments of interruptible load needs to be obtained via simulation and sensitivity prediction. Activation process of actuators is omitted. In \cite{rate_MIT}, given a disturbance, the available time remaining for resources to take actions to guarantee a bounded frequency response is estimated as a function of local inertia. This reaction time can be an appropriate metric to select an adequate response action. But again, the impact of the supportive control on this reaction time is not assessed, which must be an important factor as stronger support generally should allow longer reaction time. As seen, the second challenge is addressed by employing the system response model in analysis and control design. But for the first one, there is nearly no systematic approach to analyze the hybrid behavior.

\subsection{Contributions}
This paper proposes a systematic framework of control mode synthesis to ensure adequate frequency response. The first challenge is tackled by deriving a optimization formulation to perform reachability analysis. The second challenge is managed in a similar way to \cite{freq_limit_storage}. A composite model of SFR and WTG is used as the analytical model for reachability analysis, and as the state observer for online control. The contributions of this paper can be concluded from both theoretic and application aspects. 

From the theoretic aspect
\begin{enumerate}
	\item Based on the definition of region of safety (ROS) in \cite{zyc_hybrid_controller}, this paper formally defines the \emph{largest} ROS (LROS), and interprets the relation between the true backward reachable set and the LROS.
	\item Based on the occupation measure and corresponding formulations in \cite{ROA_Outer}, where the time-dependent backward reachable set is estimated, this paper modifies the formulation to calculate the \emph{invariant} backward reachable set, and first provides the geometric interpretation. The introduction of this theory guarantees $L$-1 norm convergence to the true LROS as a significant improvement from the framework in \cite{zyc_hybrid_controller}.
\end{enumerate}

From the application aspect
\begin{enumerate}
	\item A novel control called safety supervisory control (SSC) is proposed based on the concept of ROS. The SSC is capable of activating the grid supportive modes timely to ensure adequate system frequency response.
	\item A decentralized communication architecture is proposed for the application of the SSC in small-scale systems.
	\item A scheduling loop supplements the configuration to update the supervisor with respect to the renewable penetration levels so that the SSC is robust with respect to variations in system inertia due to stochastic and intermittent characteristics of renewables.
	\item The proposed controller is successfully implemented and verified using industrial models and commercial softwares.
\end{enumerate}

\subsection{Organization}
The rest of the paper is organized as follows. In Section \ref{sec_framework}, the optimization problem for solving the safety supervisor as well as its geometric interpretation is introduced. In Section \ref{sec_model}, the configuration of the SSC is introduced and implemented on the IEEE 39-bus system, where the dynamic performance is illustrated. Conclusions are given in Section \ref{sec_con}.\par

\section{Switching Synthesis and Supervisory Control Design via Region of Safety}\label{sec_framework}
In this section, the mode synthesis problem is defined first, followed by preliminaries on set theory and safety verification, where the concept ROS is proposed, and the safe switching synthesis principle is interpreted based on the property of ROS. Then, the main theory and formulation is expressed to explain and estimate the LROS. In Section \ref{sec_SSC_design}, the design of SSC based on the framework is described. At last, the design procedure of SSC using the proposed theory is introduced using a microgrid example. \par

\subsection{Problem Statement}
\begin{figure}[!b]
	\centering
	\includegraphics[scale=0.4]{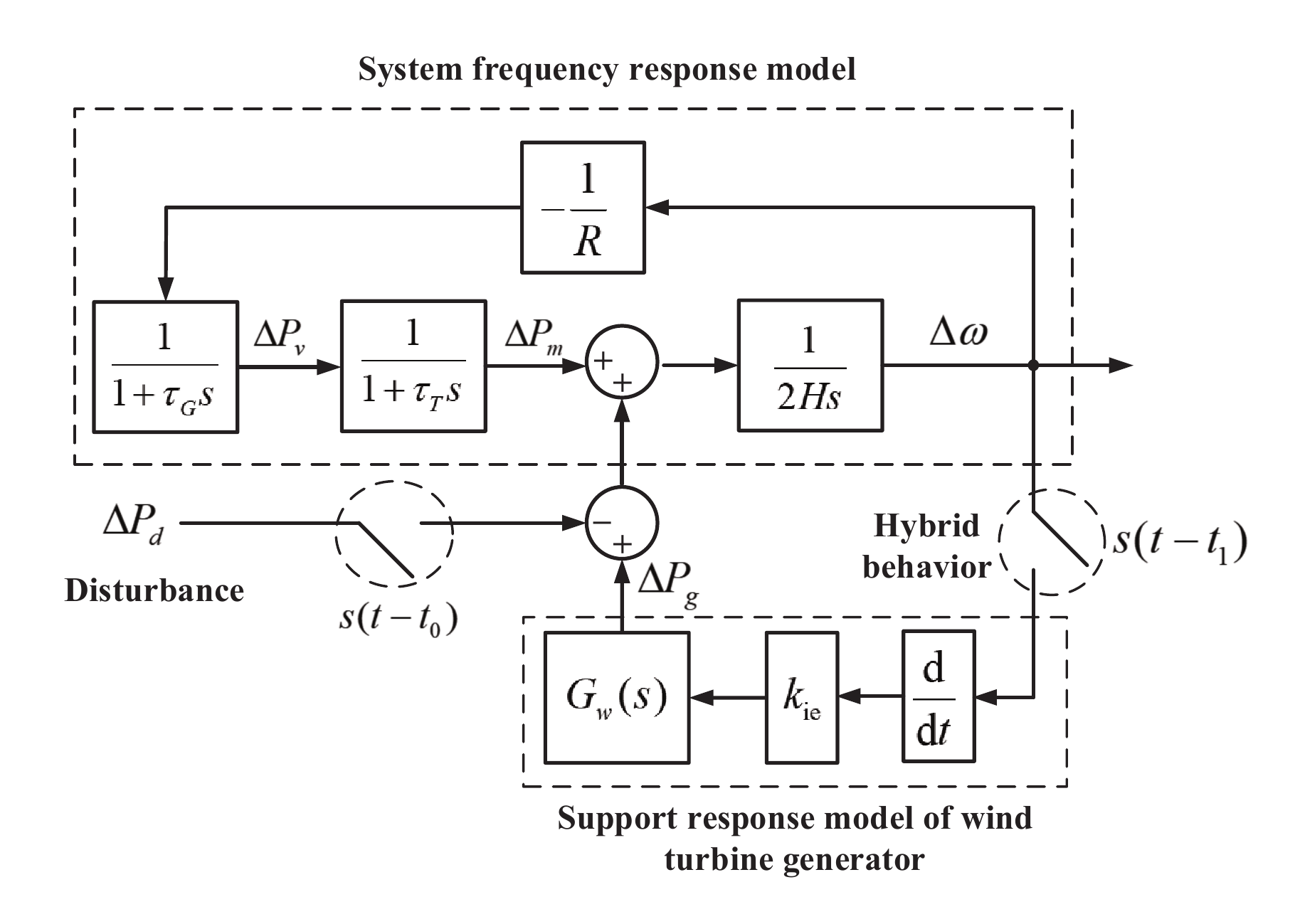}
	\caption{The hybrid system frequency response model incorporating the support response model of WTG. The supportive mode is limited to the inertia emulation for simplicity.}
	\label{fig_Concepual_Model}
\end{figure}

Consider the SFR model incorporating the support response model of WTG shown in Fig. \ref{fig_Concepual_Model}. The shifted unit step function to describe the switching behavior is given as
\begin{align}\label{eq_unit_step}
	s(t-k)=
\begin{cases}
	0 & \text{if $t<k$}\\
	1 & \text{if $t\geq k$}
\end{cases}       
\end{align}
The differential equation of the SFR can be expressed as
\begin{align}\label{eq_SFR}
\begin{aligned}
\Delta \dot{x}_{s}=A_{s}\Delta x_{s} + B_{s}k_{\text{scal}}\Delta P_{g} - B_{s}s(t-t_0)\Delta P_{d}
\end{aligned}
\end{align}
where
\begin{align*}
\begin{aligned}
&A_{s}=\left[ \begin{array}{ccc} 
0 & 0.5/H & 0\\
0 & -1/\tau_{T} & 1/\tau_{T}\\
-1/(R\tau_{G}) & 0 &  1/\tau_{G}
\end{array} \right],
B_{s}=\left[\begin{array}{c} 0.5/H\\0\\0  \end{array} \right] \\
& \Delta x_{s}=\left[\begin{array}{ccc}\Delta\omega,\Delta P_{m},\Delta P_{v}\end{array} \right]^T \\
\end{aligned}
\end{align*}
The term $\Delta P_{d}$ denotes the power imbalance due to a disturbance, which is multiplied by the shifted unit step function $s(t-t_0)$ to denote its occurrence instant $t_0$. $\Delta P_{g}$ denotes the output from the WTG associated with the grid supportive controller. The generator speed, mechanical power, valve position and governor droop are denoted by $\omega$, $P_{m}$, $P_{v}$ and $R$, respectively. The terms $H$, $\tau_{T}$ and $\tau_{G}$ denote inertia constant, turbine and governor time constant, respectively. The term $k_{\text{scal}}$ denotes a change of base if necessary. 

The other important piece of the models is to describe the supportive power from WTGs. In this paper, this model is denoted as the support response model, representing the input-output relation from the measurement signal to the active power variation as the block $G_{w}(s)$ shown in Fig. \ref{fig_Concepual_Model}. Here, the supportive function is limited to inertia emulation (IE) for simplicity. Let the support response model of the WTG be governed by the linear state-space model
\begin{align}\label{eq_WTG}
\begin{aligned}
&\Delta\dot{x}_{w}=A_{w}\Delta x_{w} + B_{w}s(t-t_1)k_{\text{ie}}\Delta\dot{\omega} \\
&\Delta P_{g}=C_{w}\Delta x_{w} + D_{w}s(t-t_1)k_{\text{ie}}\Delta\dot{\omega}
\end{aligned}
\end{align}
where $s(t-t_1)$ denotes that the IE mode is activated at $t_1$. $x_{w}$ denotes the states of the WTG. An typical example of such response models can be found in \cite{zyc_hybrid_controller}. The term $k_{\text{ie}}$ is the IE gain. 

The control objective is described as follows. Consider a computation domain of interest $X\subset \mathbb{R}^{n}$ within the state space, which can be associated with physical system limits. Assume a power imbalance occurs at time $t_{0}$. Given the IE mode with $k_{\text{ie}}$, the objective of the SSC is to activate the WTG supportive mode at time $t_{1}=t_{0}+t_{r}$ so that the frequency response is adequate, i.e., $\omega\in X_{S}=\{x|\omega^{-}_{\text{lim}}\leq\omega\leq\omega^{+}_{\text{lim}}\}\cap X$. The set $X_{S}$ is usually denoted as the \emph{safe set}, and its complementary set is called the \emph{unsafe set} $X_{U}=\{x|\omega>\omega^{+}_{\text{lim}}\text{ or }\omega<\omega^{-}_{\text{lim}}\}\cap X$. The frequency safety limits are usually defined for a set of contingencies, i.e.,  $\Delta P_{d}\in D=\{\delta|\delta^{-}_{\text{lim}}\leq \delta\leq \delta^{+}_{\text{lim}}\}$. As seen, the most important task is to determine the reaction time $t_{r}$ \cite{Uriarte2015}.

\subsection{Preliminaries}
In this subsection, the concept of ROS will be defined. Then, the safe switching principle equivalently regarding the reaction time $t_{r}$ will be explained. Having the sets of safe, unsafe, contingencies and computation been defined as $X_{S}$, $X_{U}$, $D$, and $X$, respectively, let $X_{I}\in X_{S}$ be the initial set, $\phi(t|x_{0},\Delta P_{d})$ be the trajectory initialized in $x_{0}\in X_{I}$ under disturbance $\Delta P_{d}$. Let the hybrid closed-loop system in Fig. \ref{fig_Concepual_Model} be expressed in the following compact form
\begin{align}\label{eq_SFR_WTG}
	\begin{aligned}
		\dot{x}=f_{t_{r}}(x,\Delta P_{d})
	\end{aligned}
\end{align}
where $x=[\Delta x_{s},\Delta x_{w}]^{T}$ and $t_{r}$ is the reaction time. Two concepts are defined as follows \cite{zyc_hybrid_controller}.
\begin{definition}[Safety]
	\label{thm_def_safety}
	Given (\ref{eq_SFR_WTG}), $x_0$, $X$, $X_{U}$ and $D$, the \emph{safety} property holds if there exists no time instant $T\geq 0$ and no piecewise constant bounded disturbance $d:[0,T]\rightarrow D$ such that $\phi(t|x_{0},\Delta P_{d})\cap X_{U}\neq\varnothing$ for any $t \in [0,T]$.
\end{definition}\par
\begin{definition}[Region of Safety]
	\label{thm_def_ros}
	A set that only initializes trajectories with the property in Definition \ref{thm_def_safety} is called a \emph{region of safety}.
\end{definition}\par

In other words, the ROS is a collection of initial condition $x_{0}$, starting from which the trajectories will stay within the safe set. Mathematically, the ROS can be expressed using the zero sublevel set of a continuous function $B(x)$. For a given initial set $X_{I}$, a pioneering work in \cite{Prajna2007a} is proposed to verify if it is a ROS. The theorem is given as follows.
\begin{theorem}
\label{thm_fundamental_barrier}
Let the system in (\ref{eq_SFR_WTG}), and the sets $X$, $X_{I}$, $X_{U}$ and $D$ be given, with $f$ continuous. If there exists a differentiable function $V:\mathbb{R}^{n}\longrightarrow \mathbb{R}$ such that	
\begin{align}\label{eq_barrier_certificate}
\begin{aligned}
V(x)\leq 0& \quad \forall x \in X_{I}\\
V(x)> 0& \quad \forall x \in X_{U}\\
\dfrac{\partial V}{\partial x}f(x,\Delta P_{d})<0& \qquad \forall (x,\Delta P_{d}) \in X\times D
\end{aligned}
\end{align}
then the safety of the system in the sense of Definition \ref{thm_def_safety} is guaranteed, and $X_{I}$ is a ROS.
\end{theorem}

\begin{figure}[!h]
	\centering
	\includegraphics[scale=0.2]{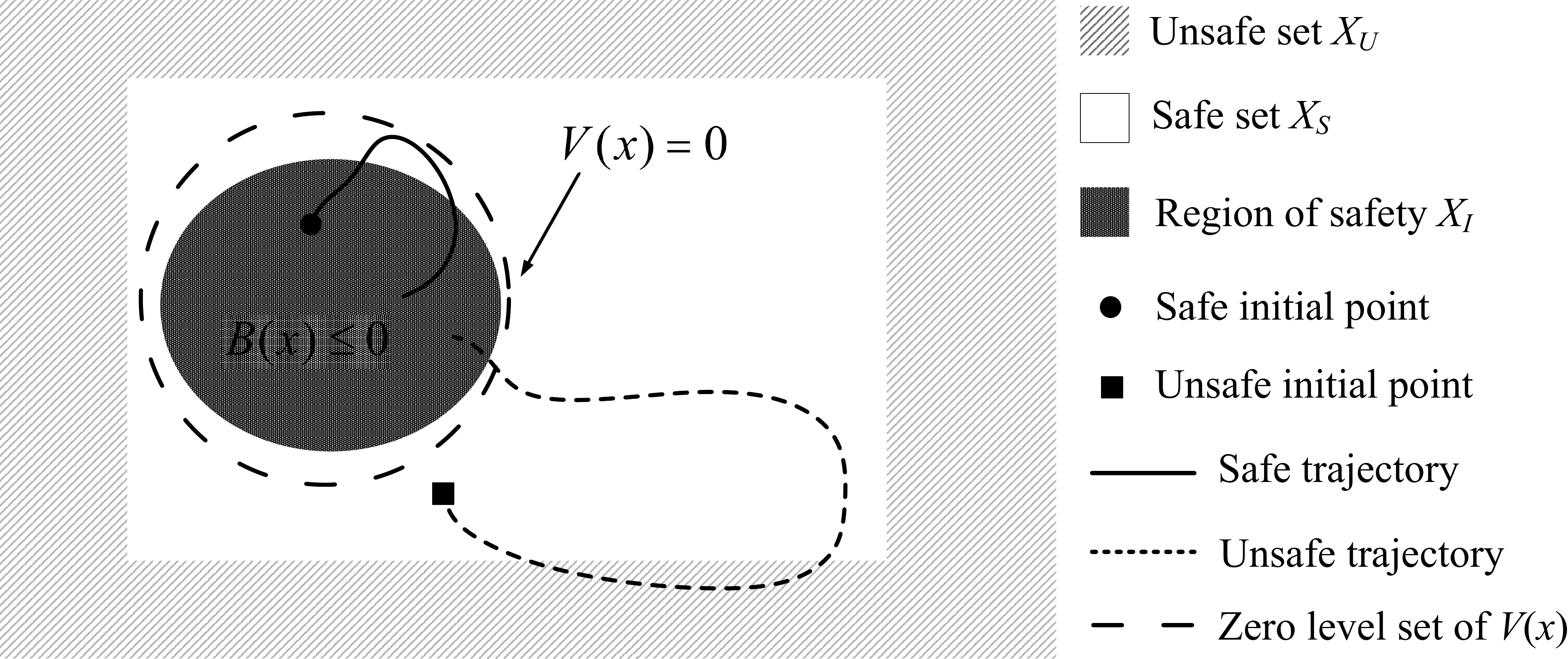}
	\caption{Illustration of the relation between the safe set, unsafe set and region of safety.}
	\label{fig_ROS_Inter}
\end{figure}

$V(x)$ is called a barrier certificate. The zero level set of $V(x)$ defines an invariant set containing $X_{I}$, that is, no trajectory starting in $X_{I}$ can leave. Thus, $X_{I}$ is a region of safety (ROS) due to the existence of $V(x)$. The relation between the safe set, unsafe set, barrier certificate and region of safety is illustrated in Fig. \ref{fig_ROS_Inter}.

\begin{figure}[!h]
	\centering
	\includegraphics[scale=0.35]{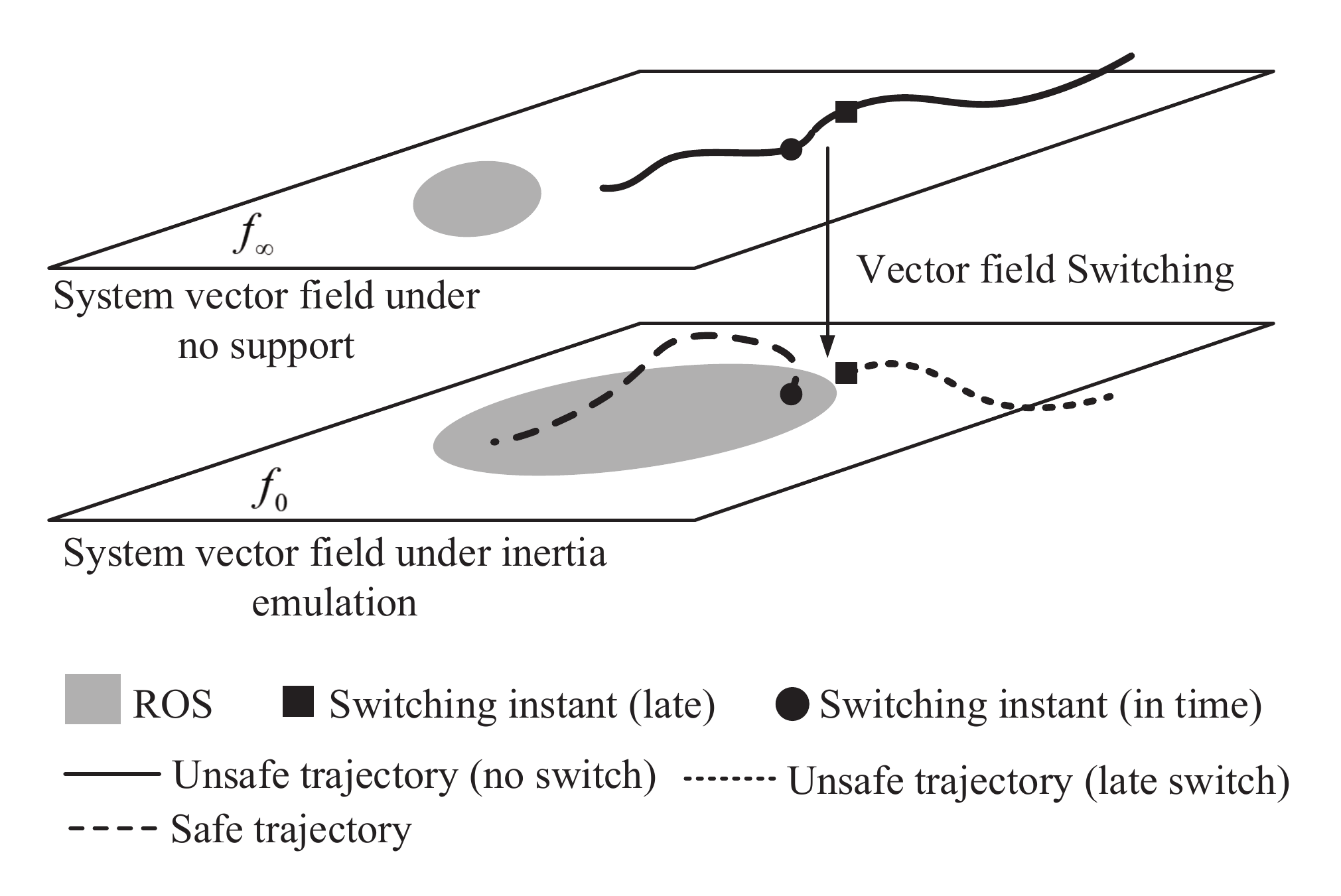}
	\caption{Switching principle under guidance of ROS for safe trajectory. The boxes are the safety limits. The green areas are the ROS of corresponding vector fields. The solid black lines are safe trajectories while the solid red ones are unsafe. The dash lines are trajectory projected onto the other vector field.}
	\label{fig_SwitchingPrinciple}
\end{figure}

Having defined these concepts, the switching synthesis principle via ROS can be interpreted as given in Fig. \ref{fig_SwitchingPrinciple}. Consider two extreme scenarios of the hybrid system in Fig. \ref{fig_Concepual_Model} when $t_{r}=\infty$ and $t_{r}=0$, respectively. The vector field $f_{\infty}(x,\Delta P_{d})$ presents the frequency response under no support, and $f_{0}(x,\Delta P_{d})$ presents the frequency response under non-delayed support. Assume the ROSs under the different vector fields are calculated for $d\in D$ and shown as the gray areas in Fig. \ref{fig_SwitchingPrinciple}. 

Due to the inertia emulation support, the ROS under its vector field is larger. When the system undergoes a contingency, a switching that guarantees adequate response can be found as long as the trajectory is inside the ROS of $f_{0}(x,\Delta P_{d})$. Since the states cannot jump, the trajectory after switching will be initialized within the ROS and according to Definition \ref{thm_def_ros} it is safe. Since the ROS is obtained in the form of a zero sublevel set of a continuous function $B(x)\leq 0$ in terms of system states, the remaining margin of a state $\widehat{x}$ to the critical switching instant can be easily found by $|B(\widehat{x})-0|$. The general principle of safe switching synthesis is subscribed by the following proposition \cite{zyc_hybrid_controller}.
\begin{proposition}\label{thm_SwitchingPrinciple}
In a hybrid system with several modes, let the ROS and trajectory of mode $i$ be denoted by $S_{i}=\{x|B_{i}(x)\leq 0\}$ and $\phi_{i}(t)$, respectively. A safe switching from mode $i$ to mode $j$ at $t_{s}$ is guaranteed if $\phi_{i}(t_{s})\in S_{j}$, that is, $B_{j}(\phi_{i}(t_{s}))\leq 0$.
\end{proposition}\par

It is clear that the key to appropriately supervising the mode switching is to estimate as close as possible the LROS, denoted by $X_{I}^{*}$. However, Theorem \ref{thm_fundamental_barrier} can only performed verification for given $X_{I}$. When $X_{I}$ is unknown, the problem becomes bilinear. Iteration approaches like one in \cite{zyc_hybrid_controller} are needed to estimate the LROS without clear information to show convergence. Thus, a more advanced technique will be introduced in the coming subsection.


\subsection{Main Theory}
\begin{figure}
	\centering
	\includegraphics[scale=0.19]{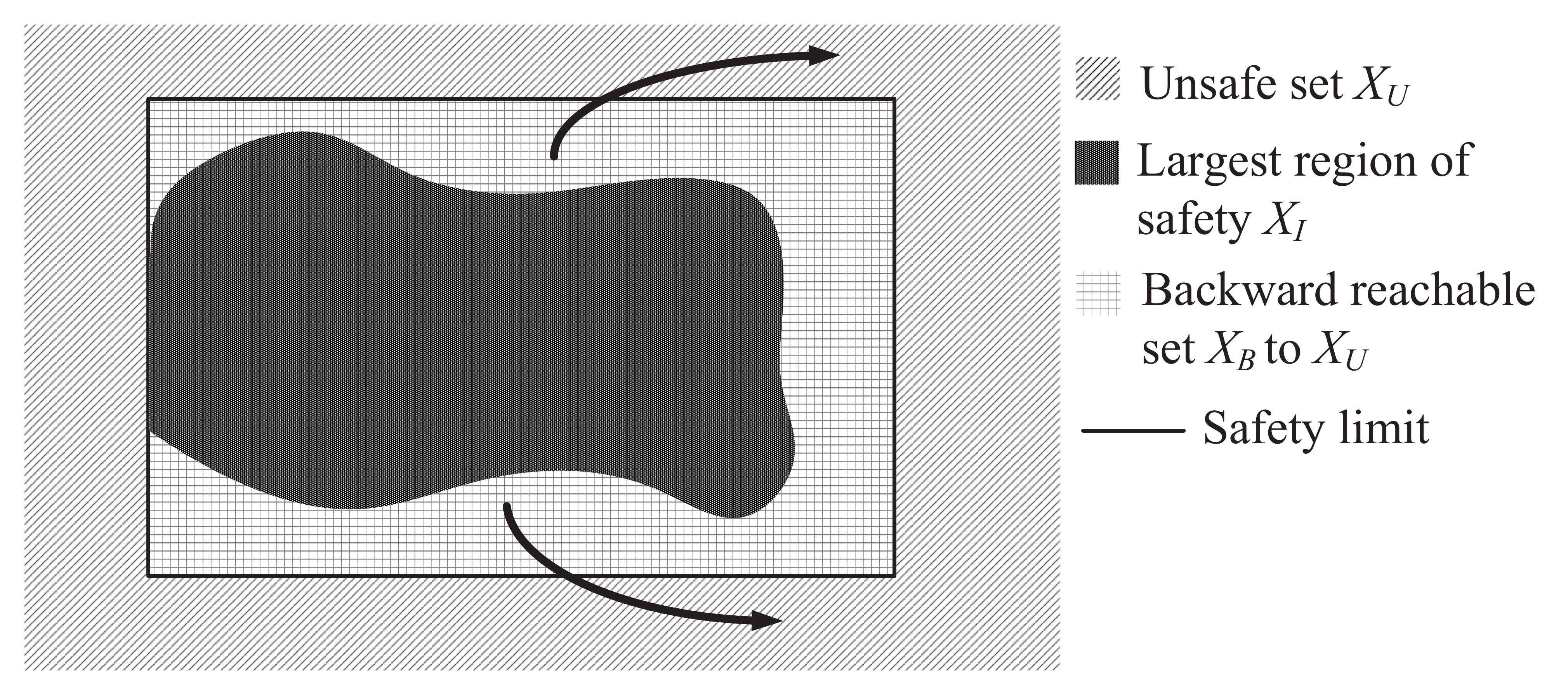}
	\caption{ROS interpretation in reachability sense. The safe set is the union of the backward reachable set to the unsafe set and the region of safety.}
	\label{fig_BackwardReach}
\end{figure}

In this subsection, the LROS will be explained in the reachability sense. Then, a formulation is proposed to estimate closely the LROS, followed by the interpretation from computational geometric point of view. Consider a computation domain of interest $X$ consisting of the safe set $X_{S}$ and unsafe set $X_{U}$ as illustrated in Fig. \ref{fig_BackwardReach}. The true and estimated invariant backward reachable set (BRS) of $X_{U}$ is denoted as $X_{B}^{*}$ and $\overline{X}_{B}$, respectively. Every trajectory starting in $X_{B}^{*}$ will reach the unsafe set. Thus, the LROS is the relative complement of $X_{B}^{*}$ with respect to $X_{S}$, i.e., $X_{I}^{*}=X_{S}\setminus X_{B}^{*}$. The BRS can be estimated by either solving the Hamilton–-Jacobi partial differential equation (PDE) \cite{hs_verify_survey}\cite{mitchell2005time} or operating sets in the form of ellipsoids \cite{renewable_variability} or zonotopes \cite{safety_2_cato,ieee_14_ac,ieee_14_ps}. The Hamilton–-Jacobi PDE approach has good convergence characteristics, but suffers from a heavy computational burden and does not scale to higher order systems. The set operation methods can be applied to more general systems due to the choice of special representations of sets, but this leads to over-approximation. 
\par

A recent novel approach proposed in \cite{ROA_Inner,korda2014convex,ROA_Outer} uses occupation measures to formulate the BRS computation as an infinite-dimensional linear program. Its dual problem is formulated on the space of nonnegative continuous functions. In \cite{ROA_Outer}, the time-dependent BRS is computed. Here, we propose the optimization problem (\ref{eq_main}) to calculate the \emph{invariant} BRS, and thus ROS, under the vector field $t_{r}=0$
\begin{subequations}
	\label{eq_main}
\begin{align}
	&\inf_{B(x),\varOmega(x)} & &\int\limits_{X}\varOmega(x)d\lambda(x)\label{eq_main_1}\\
	&\text{s.t.} & & B(x)> 0\quad\forall x\in X_{U}\label{eq_main_2}\\
	& & &\dfrac{\partial B}{\partial x}f_{0}(x,d)\leq 0\quad\forall(x,d)\in X\times D\label{eq_main_3}\\
	& & &\varOmega(x)\geq B(x)+1\quad\forall x\in X\label{eq_main_4}\\
	& & & \varOmega(x)\geq 0\quad\forall x\in X\label{eq_main_5}
\end{align}
\end{subequations}
where the computation domain $X$, unsafe set $X_{U}$ and disturbance set $D$ are given. The infimum is over $B\in C^{1}(X)$ and $\varOmega\in C(X)$. $\lambda$ denotes the Lebesgue measure. If the problem is feasible, the safety $f_{0}(x,d)$ with $d\in D$ is preserved and the zero level set of $\varOmega(x)-1$ converges below to $X_{I}^{*}$.\par

\begin{figure}[h]
	\centering
	\includegraphics[scale=0.5]{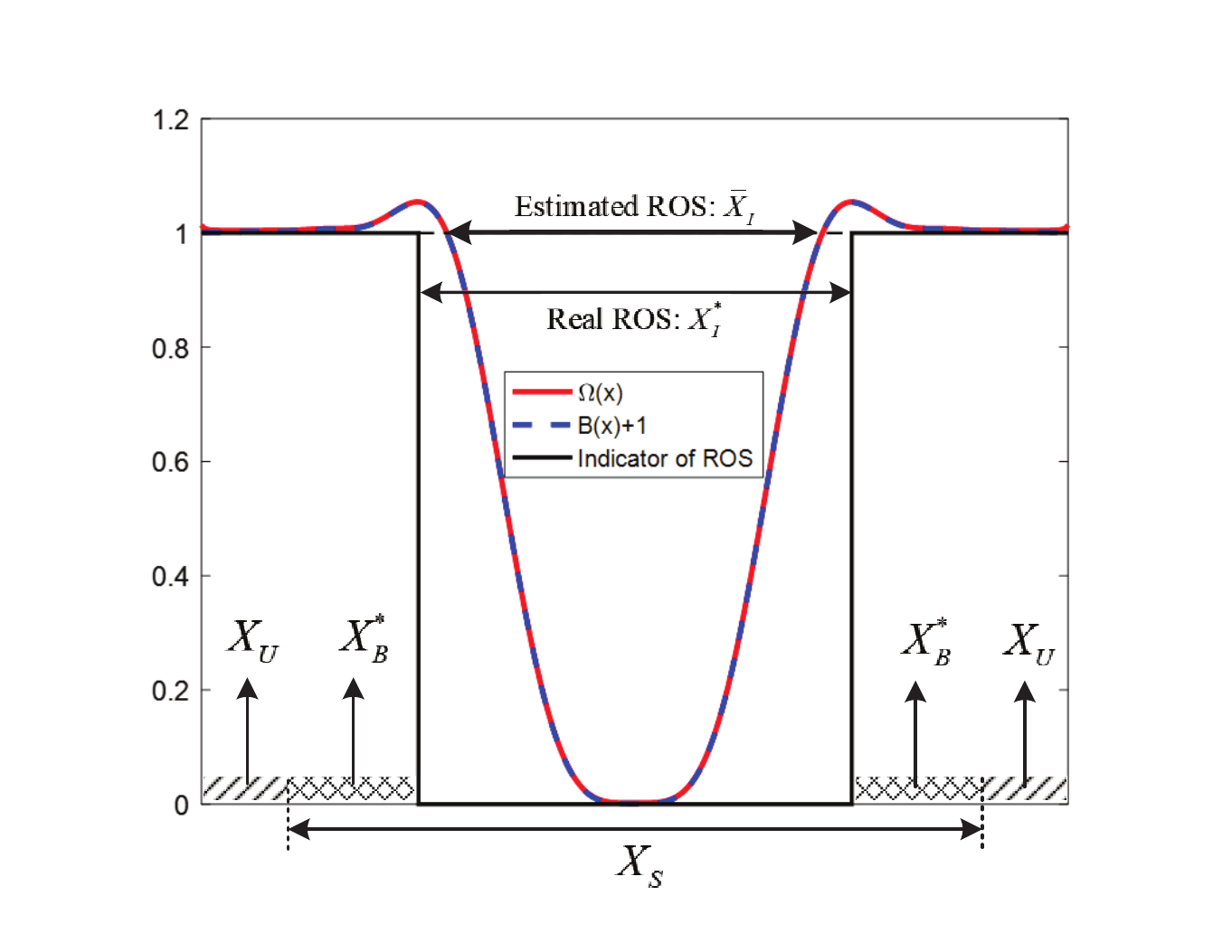}
	\caption{Geometry interpretation of proposed optimization problem for estimating the ROS. $\Omega(x)$ and $B(x)+1$ are guaranteed to be positive on $X_{U}$ and $X_{B}^{*}$.}
	\label{fig_Geometry}
\end{figure}

A strict mathematical proof is out of the scope of this paper. Instead, a geometric interpretation is given. In essence, the problem in (\ref{eq_main}) tries to estimate the BRS in Fig. \ref{fig_BackwardReach} without knowing the initial set $X_{0}$. Let any trajectory eventually ending up in the set $X_{U}$ at certain time $T$ be denoted as $\phi(T|x_{0})$. Based on the conditions of $B(\phi(T|x_{0}))>0$ in (\ref{eq_main_2}) and the passivity in (\ref{eq_main_3}), one can easily show $B(x_{0})>0$. Thus, (\ref{eq_main_2}) and (\ref{eq_main_3}) ensure that $B(x)>0$ for any $x\in X_{B}^{*}$ illustrated as a one dimensional case in Fig. \ref{fig_Geometry}. The conservatism lies in the fact that $B(x)>0$ for some $x\in X_{I}^{*}$, which overestimates the BRS (i.e., ${X}_{B}^{*}\subset\overline{X}_{B}$) and in turn underestimates the ROS (i.e., ${X}_{I}^{*}\supset\overline{X}_{I}$). Fortunately, this conservatism can be reduced by introducing a positive slack function $\varOmega(x)$ that is point-wise above the function $B(x)+1$ over the computation domain $X$. Assume the complement set of $X_{I}^{*}$ is represented by the indicator function $\delta_{X\setminus X_{I}^{*}}(x)$, i.e., a function is equal to one on $X\setminus X_{I}^{*}$ and 0 elsewhere. The key idea of the problem in (\ref{eq_main}) is that by minimizing the area of function $\varOmega(x)$ over the computation domain $X$, the function $B(x)+1$ will be forced to approach $\delta_{X\setminus X_{I}^{*}}(x)$ from above as shown in Fig. \ref{fig_Geometry}. Thus, the zero sublevel set of $\varOmega(x)-1$ is an inner approximation of $X_{I}^{*}$. Essentially, the problem in (\ref{eq_main}) is trying to approximate an indicator function using a polynomial. The conservatism of the estimate vanishes with increasing order of the polynomial.\par

Obviously the problem in (\ref{eq_main}) attains its optimum when $\delta_{X\setminus X_{I}^{*}}(x)=B(x)+1=\varOmega(x)$. If all functions are confined to be polynomials and all sets are basic closed semi-algebraic sets\footnote{According to the Weierstrass approximation theorem in real analysis, every continuous function defined on a closed interval can be uniformly approximated as closely as desired by a polynomial function. Since almost every dynamic equation in power system is continuous, limiting class of functions to be polynomial is sufficient in describing the system characteristics, and meanwhile provides algorithmic feasibility.} (hence defined by finitely many polynomial inequalities and equality constraints) \cite{siam_rev}, then the recent SOS decomposition techniques \cite{sosd} can reformulate problem (\ref{eq_main}) into a SOS program, which can be further converted into a semi-definite program (SDP). This procedure has been implemented in several toolboxes such as Yalmip \cite{sos_yalmip}. The SOS program to solve problem (\ref{eq_main}) is given as follows.\par
Let $X=\left\lbrace x\in\mathbb{R}^{n}|g_{X}(x)\geq 0\right\rbrace $, $X_{U}=\left\lbrace x \in \mathbb{R}^{n}|g_{U}(x)\geq 0\right\rbrace $, and $D=\left\lbrace d \in \mathbb{R}^{m}|g_{D}(d)\geq 0\right\rbrace$, which are represented by the zero superlevel set of the polynomials $g_{X}(x)$, $g_{U}(x)$, and $g_{D}(d)$, respectively, and some small positive number $\epsilon$ be given. Functions $B(x)$ and $\varOmega(x)$ are polynomials with fixed degree. Multipliers $\sigma_{i}(x)$ for $i=1,\cdots,6$ are SOS polynomials with fixed degree. Then the ROS can be obtained by solving the following SOS program
\begin{subequations}\label{eq_sos}
\begin{align}
\inf_{B(x),\varOmega(x)} & \omega'l\\
B(x)-\epsilon - \sigma_{1}(x)g_{U}(x)& \in \varSigma^{2}\left[ x\right]\\
\begin{split}
-\dfrac{\partial B}{\partial x}(x)f_{0}(x,d)-\sigma_{2}(x,d)g_{D}(d)\\-\sigma_{3}(x,d)g_{X}(x)&\in \varSigma^{2}\left[ x\right]\label{eq_sos_3}
\end{split}\\
\varOmega(x)-B(x)-1-\sigma_{4}(x)g_{X}(x) &\in \varSigma^{2}\left[ x\right]\\
\varOmega(x)-\sigma_{5}(x)g_{X}(x) &\in \varSigma^{2}\left[ x\right]
\end{align}
\end{subequations}
where $l$ is the vector of the moments of the Lebesgue measure over $X$ indexed in the same basis in which the polynomial $\varOmega(x)$ with coefficients $\omega$ is expressed. For example, for a two-dimensional case, if $\varOmega(x)=c_{1}x_{1}^{2}+c_{2}x_{1}x_{2}+c_{3}x_{2}^{2}$, then $\omega=[c_1,c_2,c_3]$ and $l=\int_{X}[x_{1}^{2},x_{1}x_{2},x_{2}^{2}]\text{d}x_{1}\text{d}x_{2}$.

\subsection{Design Procedure of Safety Supervisory Control for Single WTG}\label{sec_SSC_design}
\begin{figure}[h]
	\centering
	\includegraphics[scale=0.48]{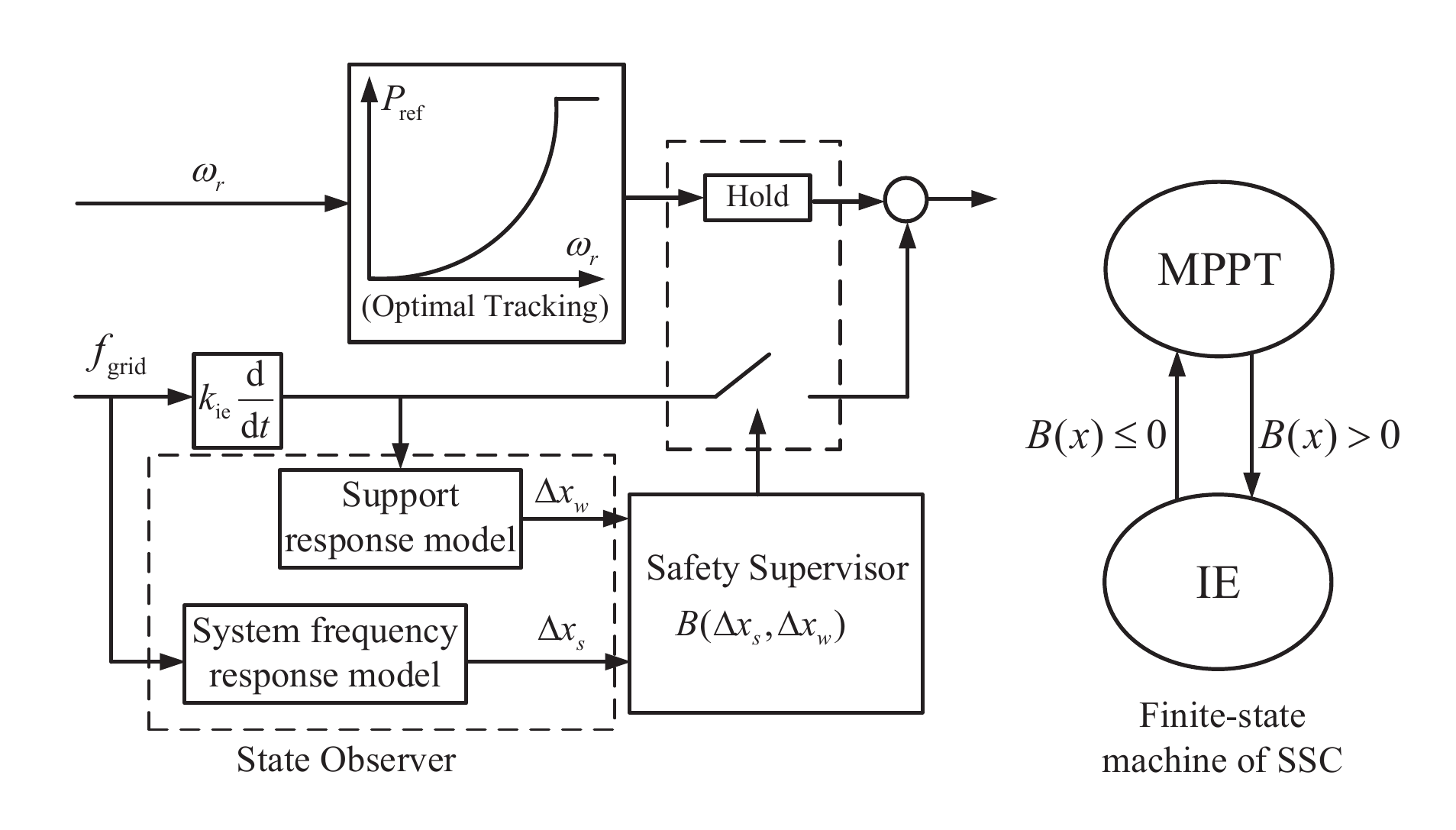}
	\caption{Safety supervisory control (SSC) integrated in WTGs and its corresponding finite-state machine. The SSC enables the system awareness capability and provides real-time systemic safety margin for WTGs.}
	\label{fig_SSC_Config}
\end{figure}

Based on the Prop. \ref{thm_SwitchingPrinciple}, the real-time margin for safe switching can be obtained by checking the current level set of states in the ROS of the IE mode. A safe switching can be committed before the level set of states in the ROS becomes positive. As analyzable and quantifiable, the ROS is deployed online as a replacement of deadbands to switch the modes of a WTG. Meanwhile, a state observer is equipped to provide inputs to the ROS. The integrated system is denoted as the \emph{safety supervisory control}. The configuration and the finite-state machine of the WTG is given in Fig. \ref{fig_SSC_Config}. The design procedure is given as follows
\begin{enumerate}
	\item Build the SFR model incorporating the support response model of WTG.
	\item Compute the ROS of the SFR under non-delayed support of WTG.
	\item Deploy the ROS online as the safety supervisor.
	\item Estimate the states of the grid using the SFR model with frequency input, and estimate the states of the WTG using the support response model with IE signal.
\end{enumerate}

\subsection{Illustrative Example: a Diesel/Wind fed Microgrid}\label{sec_MG_example}
\begin{figure}[h]
	\centering
	\includegraphics[scale=0.4]{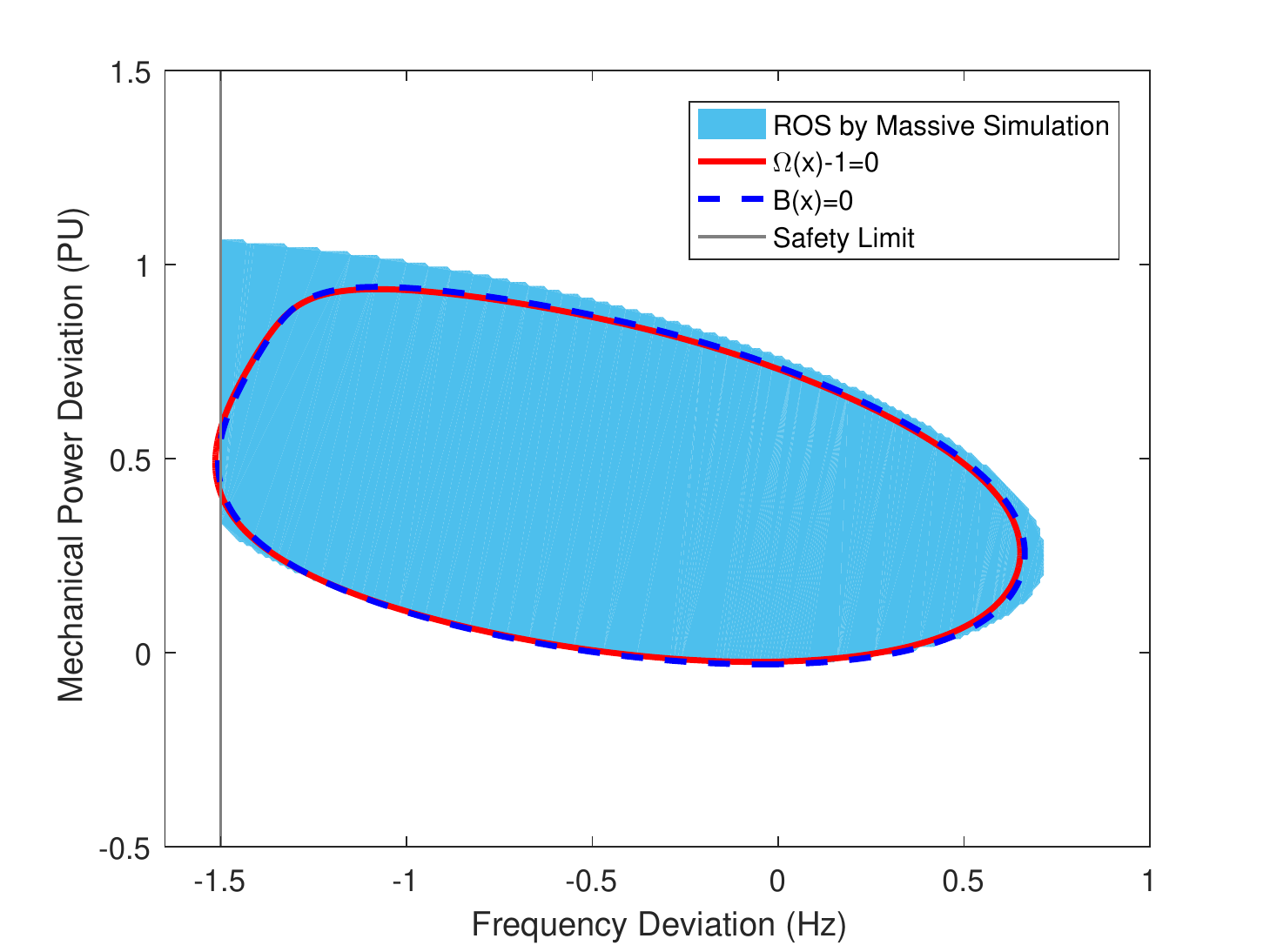}
	\caption{Comparison of ROS (projected onto $\Delta\omega-\Delta P_{m}$ plane) between the subzero level set of $B(x)$ (blue dash) and exhaustive simulations (blue).}
	\label{fig_ROS}
\end{figure}

To illustrate the SSC, a lumped diesel/wind fed microgrid in \cite{zyc_ISGT2017} is investigated. Assume that the parameters of the frequency response model of the diesel generator in the form of (\ref{eq_SFR}) have been estimated. The WTG model used in this section is a first-order linear system obtained using the selective modal analysis based model reduction \cite{hector}. Then, the rotor speed of WTG is the state, i.e., $\Delta x_{w}=\Delta\omega_{r}$. The parameters are given as follows
\begin{equation*}
\begin{split}
& H=2,R^{-1}=30,\tau_{G}=0.1,\tau_{T}=0.5\\
& A_{w}=-0.3914,B_{w}=-0.3121,C_{w}=1.37,D_{w}=1\\
& k_{\text{scal}}=0.15, k_{\text{ie}}=0.2,\Delta P_{d}\in D=\{d|0\leq d\leq 0.32\}
\end{split}
\end{equation*}
and the states of the overall frequency response model including synthetic inertial response are $x=[\Delta\omega,\Delta P_{m},\Delta P_{v},\Delta\omega_{r}]$.
The safety limit is set as $\omega^{-}_{\text{lim}}=58.5$ Hz. With all the given conditions, the problem in (\ref{eq_sos}) is formulated in Yalmip \cite{yalmip} and solved by Mosek \cite{mosek}. The ROS is represented by the zero sublevel set of $B(x)$ and its projection on the phase plane of frequency and mechanical power is shown in Fig. \ref{fig_ROS}. The blue region is the ROS obtained by massive simulations and can be considered as the "true" region. As shown by minimizing the area under the slack function $\varOmega(x)$, the zero level set of $B(x)$ is expanded by $\varOmega(x)-1$ as much as possible to the true LROS under a fixed highest degree.\par

Once $B(x)$ is obtained, it will be deployed online in the configuration shown in Fig. \ref{fig_SSC_Config}. The speeds of diesel and wind turbine generators are measurable. The $\Delta P_{m}$ and $\Delta P_{v}$ can be estimated using the SFR model. The SSC integrated into WTGs not only enables the situational awareness capability, but also provides a real-time margin towards safe switching.\par

\begin{figure}[h]
	\centering
	\includegraphics[scale=0.4]{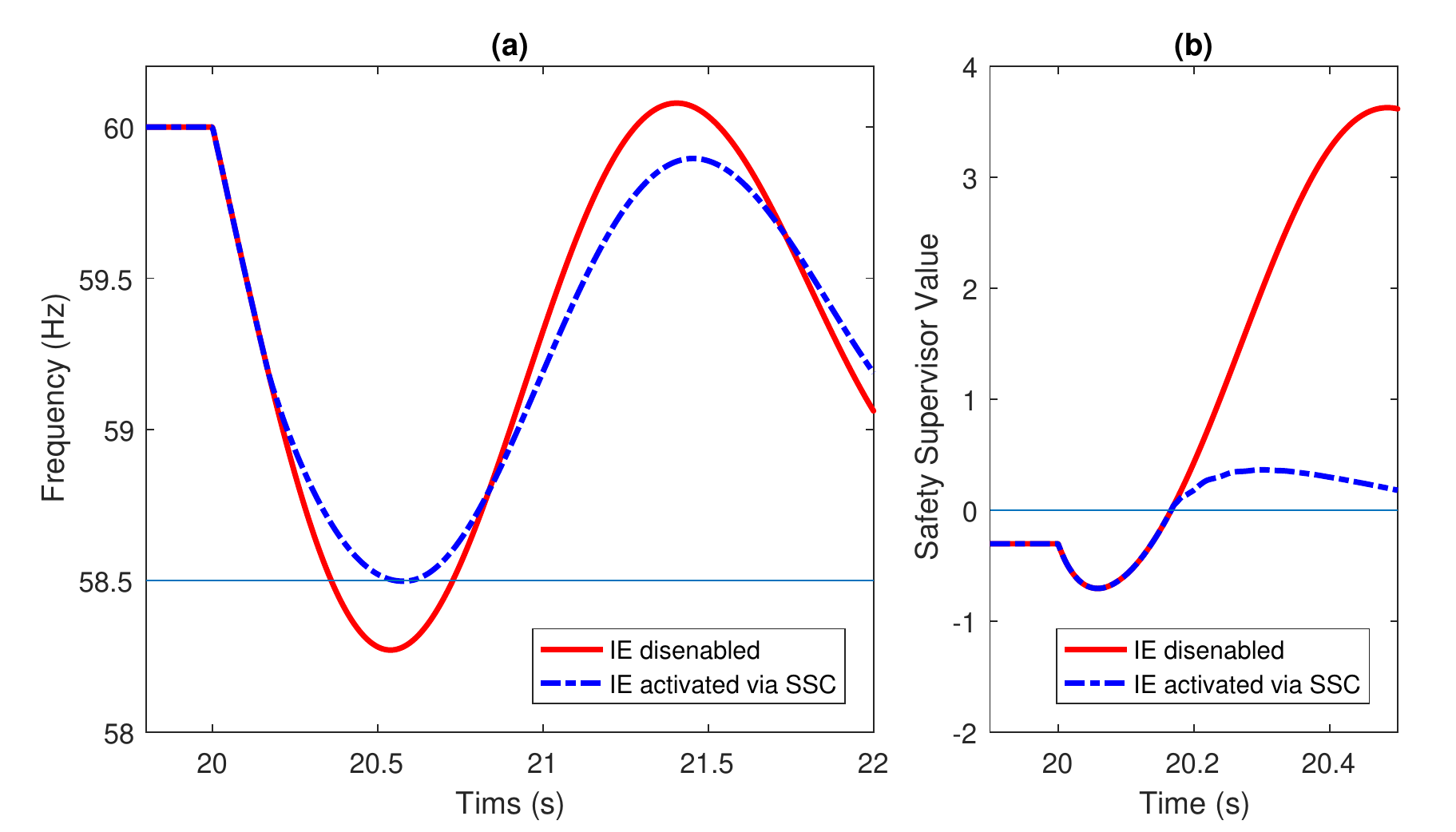}
	\caption{Frequency response under no inertia emulation and inertia emulation activated via safety supervisory control (SSC), (a) frequency response, (b) value of safety supervisor.}
	\label{fig_ROS1}
\end{figure}

To show in a simulation, the full-order nonlinear model of a synchronous generator (SG) is used but scaled down to microgrid rating. A type-4 wind turbine with an averaged converter model is used. Detailed description of model used in simulation can be found in \cite{zyc_ISGT2017}. 
The system under the worse-case disturbance is simulated. The frequency response and the value of safety supervisor are shown in Fig. \ref{fig_ROS1}. The IE is activated when the supervisor's value crosses zero. As seen, the nadir of the frequency response with activated SSC is exactly at the safety limit, indicating the estimated LROS is highly precise.

\section{Decentralized Safety Supervisory Control for Multi-machine Systems}\label{sec_model}
\subsection{Center of Inertia Frequency and System Frequency Response Model}\label{sec_model_a}
In Section \ref{sec_MG_example}, design procedures of the SSC for a single-machine system have been demonstrated. The proposed framework ensures accurate estimation of the LROS. Thus, the key for adequate system frequency response is to build a precise SFR model including the synthetic inertial response from WTGs. Due to the increasing computational complexity of SOS decomposition with respect to system dimensions, the multi-machine models will make the problem computationally intractable even based on the state-of-the-art computation capability. Thus, the COI frequency response model is adopted in this paper. The COI frequency, or frequency at the equivalent inertial center, has been widely used in system frequency behavior \cite{GE_SFR, NREL_EI_SFR, Lawrence_SFR}. The frequency deviation of a single machine (area) from the COI frequency is determined by the electrical distance to the inertial center, which is further determined by the line impedance \cite{anderson1990low}. Based on this deviation, extra margins can be added to the safety limit to prevent the frequency of any single machine (area) from reaching the under-frequency load shedding (UFLS) zone.  \par

It is very difficult, however, to determine this margin theoretically and optimally. A detailed transmission line and wave propagation model may be needed to perform necessary analysis, which is of scope of this paper. On the other hand, arbitrary setting of the margin will confuse the demonstration on accuracy. Thus, we intentionally leave no extra margin in the following demonstration. Under this setting, individual SG frequencies will exceed the safety limit for a short period of time. This duration is determined by the distance between individual frequencies and the COI frequency, which is further determined by the electric distance, that is, system-dependent (since the SSC is designed towards a worst case). For systems small enough, guaranteeing the safety of COI frequency can significantly reduce the possibility of transient under-frequency relay action. 

Let $\mathcal{S}$ denote the index set of synchronous generators. Let $\mathcal{W}$ denote the index set of WTGs that have been selected as actuators of SSC. $N_{s}$ and $N_{w}$ denote the total number of generators in each set, respectively. The model in (\ref{eq_SFR}) will serve as the COI frequency response model, except that the COI inertia constant $H_{\text{coi}}$ is calculated as
\begin{align}\label{eq_COI}
\begin{aligned}
H_{\text{coi}}=\dfrac{\sum_{i\in \mathcal{S}}^{N_{s}}S_{i}^{s}H_{i}^{s}}{S_{\text{sg}}},S_{\text{sg}}=\sum_{i\in \mathcal{S}}^{N_{s}}S_{i}^{s}
\end{aligned}
\end{align}
where $S_{i}^{s}$ and $H_{i}^{s}$ are the base and inertia constant of synchronous generator $i$, respectively. The governor and turbine models represent the averaged mechanical behavior of the overall system. It is assumed that the corresponding time constants have been estimated.\par

\begin{figure}[h]
	\centering
	\includegraphics[scale=0.4]{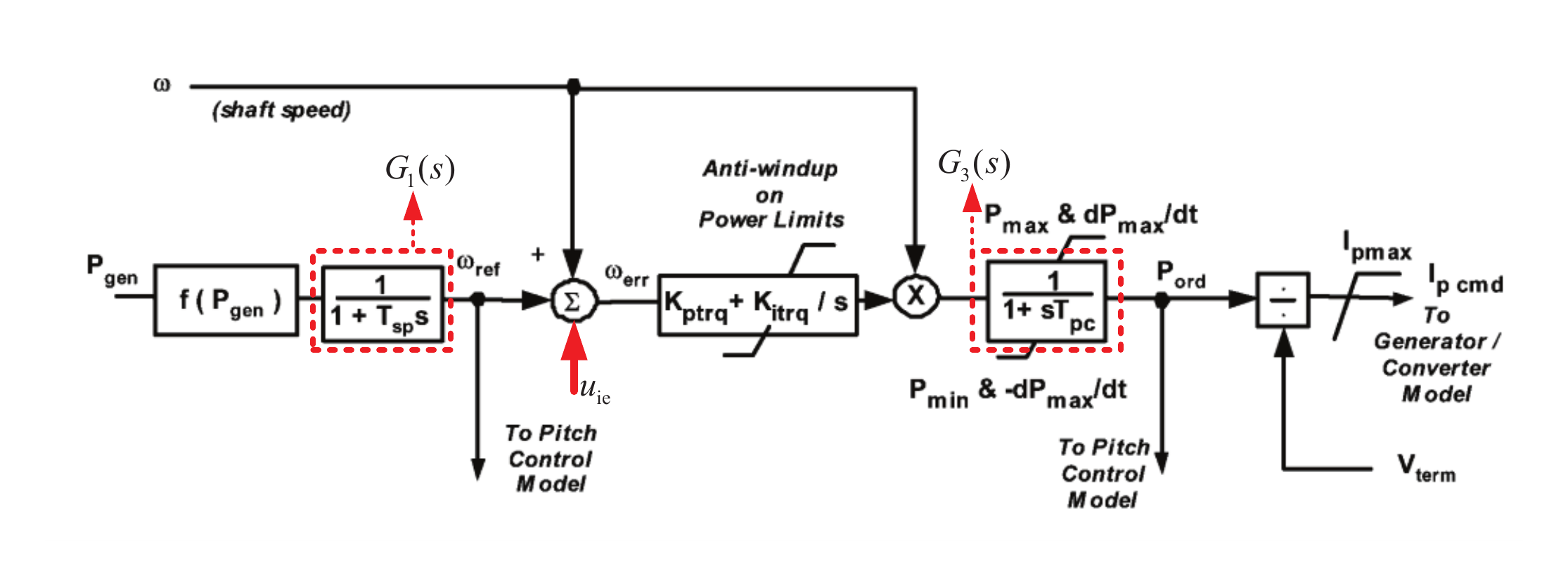}
	\caption{Widely used active power control loop for western electricity coordinating council generic type-3 wind turbine generator model \cite{hiskens,ge_report,DSA_UDM}.}
	\label{fig_ActivePowerControl}
\end{figure}

The western electricity coordinating council (WECC) generic type-3 WTG model and corresponding controls detailed in \cite{hiskens} is used. The active power control loop is shown in Fig. \ref{fig_ActivePowerControl}. The low-pass filter $G_{1}(s)$ aims to filter out the fluctuation from the MPPT signal, where its time constant $T_{\text{sp}}$ is usually in the time frame of tens of seconds \cite{ge_report}. During the inertial and primary frequency response, the reference signal $\omega_{\text{ref}}$ can be assumed constant. The transfer function $G_{3}(s)$ models the inner current loop dynamics of converter controllers. As the current regulation is in the time frame of milliseconds, this part can be omitted \cite{slootweg2003general}.
Similar to the SFR model, an aggregated model will represent the overall behavior of WTGs under supportive modes. Based on the above simplifications, the WTG responsive model associated with the IE mode is
\begin{align}\label{eq_gen_WTG}
\begin{aligned}
\dot{\overline{x}}&=\overline{K}_{\text{itrq}}(\overline{\omega}_{r}-\overline{\omega}_{\text{ref}} + \overline{u}_{\text{ie}})\\
\dot{\overline{\omega}}_{r}&=\dfrac{1}{2\overline{H}_{w}\overline{\omega}_{r}}(\overline{P}_{m,w}-\overline{\omega}_{r}\overline{y})
\end{aligned}
\end{align}
where
\begin{align}
\begin{aligned}
\overline{y}&=\overline{x}+\overline{K}_{\text{ptrq}}(\overline{\omega}_{r}-\overline{\omega}_{\text{ref}} + \overline{u}_{\text{ie}})\\
\overline{P}_{G}&=\overline{\omega}_{r}\overline{y}
\end{aligned}
\end{align}
and the averaged inertia constant of WTGs is calculated as
\begin{align}\label{eq_COI_WTG}
\begin{aligned}
\overline{H}_{w}=\dfrac{\sum_{i\in \mathcal{W}}^{N_{w}}S_{i}^{w}H_{i}^{w}}{S_{\text{wt}}},S_{\text{wt}}=\sum_{i\in \mathcal{W}}^{N_{w}}S_{i}^{w}
\end{aligned}
\end{align}
where $S_{i}^{w}$ and $H_{i}^{w}$ are the base and inertia constant of WTG $i$, respectively. In (\ref{eq_gen_WTG}), $\overline{u}_{\text{ie}}$ generated from the COI frequency is the input and the power variation $\overline{P}_{G}$ with the base of $S_{\text{wt}}$ is the output. The aerodynamic model in \cite{DSA_UDM} is employed, where $\overline{P}_{m,w}$ is a function of $\overline{\omega}_{r}$, wind speed and pitch angle. Under the time snapshot of inertial and primary response, wind speed, pitch angle and $\overline{\omega}_{\text{ref}}$ are assumed to be fixed. As shown in \cite{zyc_hybrid_controller}, linearized models are able to capture the input-output relation from the rate of change of frequency to the supportive power variations of WTGs. By linearizing (\ref{eq_gen_WTG}) and applying a change of base as $k_{\text{scal}}=S_{\text{wt}}/S_{\text{sg}}$, one can have the overall model in the form of (\ref{eq_SFR_WTG}) for ROS computation.

As for the gain of IE, when the grid frequency measurement (rather than predetermined surge signal) is used as the input signal, it is difficult to determine analytically an adequate gain since it depends on the interaction between WTGs and synchronous generators. Nevertheless, we proposes an approximated equation to provide guideline of the gain design
\begin{align}
\label{eq_WTG_gain}
\begin{aligned}
k_{\text{ie},i}&=\rho_{i}\frac{k_{\text{ad}}(\Delta P_{d}-0.5A_{m}T_{\text{nad}})}{S_{i}^{w}A_{r}}
\\&\leq\frac{\min\{P_{\text{max}}-p_{g,i},p_{g,i}-P_{\text{min}}\}}{S_{i}^{w}A_{r}}
\end{aligned}
\end{align}
where $T_{\text{nad}}$ is the time duration from the instant of disturbance occurrence to the one when frequency reaches its nadir, $A_{r}$ and $A_{m}$ is the averaged value of rate of change of frequency and rate of change of mechanical power during $T_{\text{nad}}$, respectively. The terms $S_{i}^{w}$, $P_{\text{max}}$, $P_{\text{min}}$ and $p_{g,i}$ represent the base, output power upper, lower limit and current output, respectively, of WTG $i$. $P_{\text{max}}$ is associated with the capacity rating, while $P_{\text{min}}$ is associated with the rotor speed security. $\Delta P_{d}$ is the disturbance. The term $(\Delta P_{d}-0.5A_{m}T_{\text{nad}})$ is an approximation of total inertial response in watt, and $k_{\text{ad}}$ is the required percentage from all coordinated WTGs to secure frequency, and is mainly impacted by the system total inertia. $\rho_{i}$ is the required contribution percentage of WTG $i$. When the right-hand side of Eq. (\ref{eq_WTG_gain}) is not binding, $\rho_{i}$ can be set to $100\%$, and only WTG $i$ is providing support. Otherwise, $\rho_{i}$ will be adjusted from $0\%$ to $100\%$ to satisfy the operating constraints of WTG $i$, and more WTGs will be coordinated such that the summation of $\rho_{i}$ will be $100\%$. Note that in Eq. (\ref{eq_WTG_gain}), $A_{m}$, $A_{r}$ and $k_{\text{ad}}$ need to be determined through a trial-and-error adjustment. The derivation of Eq. (\ref{eq_WTG_gain}) is explained in Appendix \ref{apen_gain}. As seen, once the worst-case contingency $\Delta P_{d}$ is given, the controller gain is a function of number of actuators and their current outputs.


\subsection{Decentralized Communication for Small-scale Systems}\label{sec_model_b}
\begin{figure}[h]
	\centering
	\includegraphics[scale=0.45]{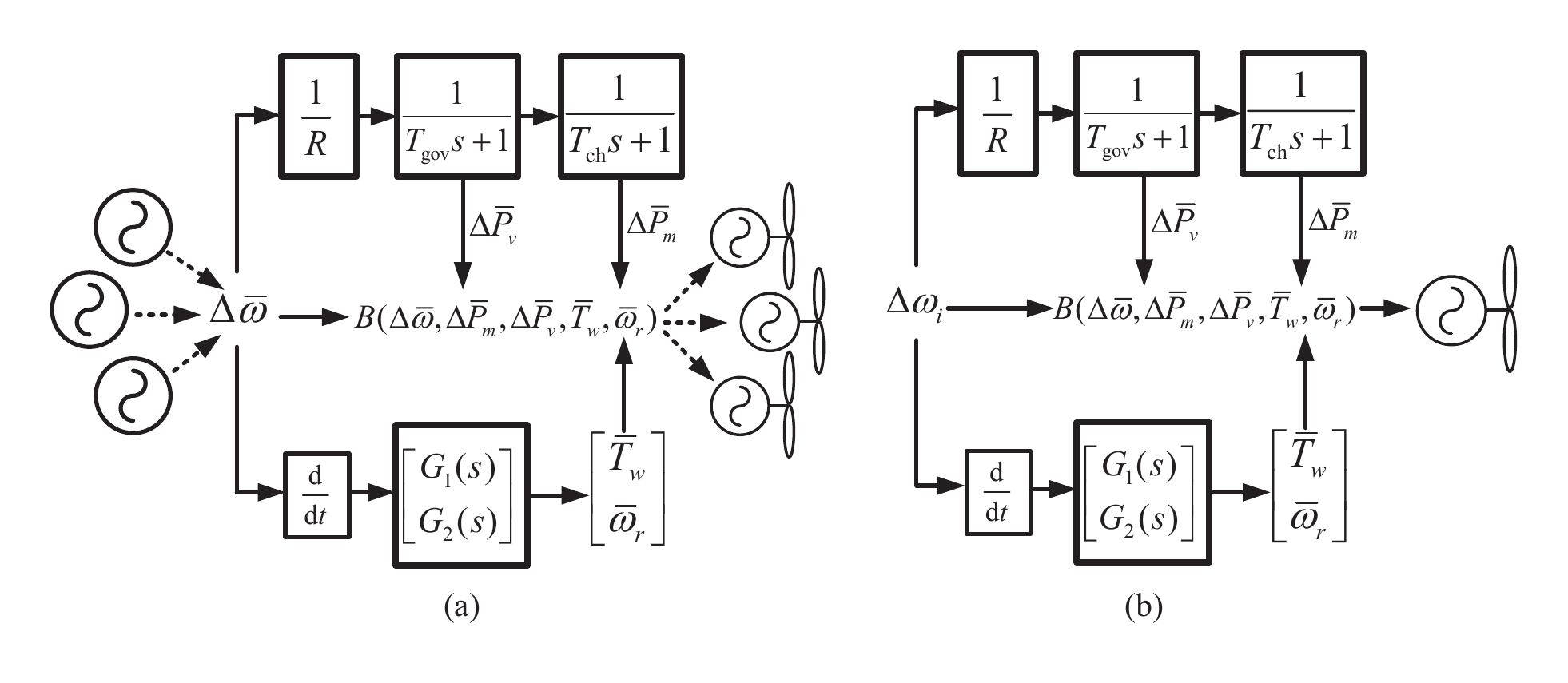}
	\caption{Centralized and decentralized communication in SSC. The decision results will be equivalent for a small scale system. (a) Centralized communication. (b) Decentralized communication.}
	\label{fig_Communication}
\end{figure}

Based on the theoretical model developed in Section \ref{sec_model_a}, a centralized communication link shown in Fig. \ref{fig_Communication} (a) is needed for the SSC. The speed of each synchronous generator (area) is measured and transmitted to the central controller to calculate the COI frequency. Then, the COI frequency is sent to state observers to estimate states $\Delta \overline{P}_{m}$, $\Delta \overline{P}_{v}$, $\Delta \overline{x}$ and $\Delta \overline{\omega}_{r}$. Finally, all the states are substituted into the safety supervisor to make switching decisions. This switching signal needs to be transmitted to each WTG to activate the inertia emulation. Although such a communication architecture could theoretically ensure the safe response of the COI frequency, it would likely introduce excessive delay and complexity, reduce reliability and require addtional infrastructure cost.\par

Essentially, the WTGs only need the switching signal, which is determined by predicting overall system behavior. Since frequency is a global feature, the system awareness capability can be integrated locally in each WTG using the same state observer. Then, as long as the input is the COI frequency, the result will be the same. In order to further reduce the communication links, measuring local frequency is desired. It is known that the local frequency will deviate from the COI frequency during transient period. But for a small-scale system, such deviations should be small. Thus, it is reasonable to assume that the frequency of single machine (area) approximates the COI frequency, i.e., $\omega_{i}\approx\overline{\omega}$. Therefore, the centralized SSC can be replaced by a decentralized SSC as shown in Fig. \ref{fig_Communication} (b). The decentralized SSC is completely integrated into a single WTG, and only local frequency measure are needed. Still, such communication reduction is only equivalent when the system is small.

\subsection{Simulation to Verify Dynamic Performance}\label{sec_model_c}
In this subsection, the modified New England IEEE 39-bus system with more than 50\% wind generation is used to demonstrate the SSC. Two scenarios with varying levels of wind generation used as actuators and different inertia emulation gains are illustrated.

The modified generator data of the system is listed in Table \ref{tab_39_bus}. The bold inertia constants indicate that they are visible to the grid. 
\begin{table}[htbp!]
	\caption{Generator Data of Modified New England 39-bus System}
	\centering
	\begin{tabular}{lclclclclclclclcl}
		\toprule 
		\# & Bus & Type & Output (MW) & Base (MVA) & Inertia (s) \\
		\midrule
		1 & 30 & WTG & 550 &$S_{1}=670$	&$H_{1}=8.00$ \\
		2 & 31 & WTG & 572 &$S_{2}=670$	&$H_{2}=8.00$\\
		3 & 32 & WTG & 650 &$S_{3}=670$	&$H_{3}=8.00$\\
		4 & 33 & SG & 632  &$S_{4}=1000$&$H_{4}=\textbf{2.86}$\\
		5 & 34 & WTG & 508 &$S_{5}=670$	&$H_{5}=8.00$\\ 
		6 & 35 & WTG & 650 &$S_{6}=670$	&$H_{6}=8.00$\\
		7 & 36 & SG & 400  &$S_{7}=1000$&$H_{7}=\textbf{2.64}$\\
		8 & 37 & WTG & 540 &$S_{8}=670$	&$H_{8}=8.00$\\
		9 & 38 & SG & 830  &$S_{9}=1000$&$H_{9}=\textbf{3.45}$\\
		10 & 39 & SG & 859 &$S_{10}=1000$&$H_{10}=\textbf{5.00}$\\
		\bottomrule
	\end{tabular}
	\label{tab_39_bus}
\end{table}
The traditional plant pool is $\mathcal{S}=\{4,7,9,10\}$. The synchronous generators are round rotor models equipped with 1992 IEEE type DC1A excitation system model (ESDC1A) and steam turbine-governor model (TGOV1) \cite{siemens2009pss}. The aggregation procedure for parameters in TGOV1 can be found in \cite{Shi2018Analytical} and given as follows 
\begin{equation*}
\begin{split}
& \overline{R}=0.05,\overline{T}_{1}=0.5,\overline{T}_{2}=2,\overline{T}_{3}=6,\overline{D}_{t}=0
\end{split}
\end{equation*}
The Western Electricity Coordinating Council (WECC) generic type-3 WTG model with built-in controls in \cite{DSA_UDM}, which includes the active power control loop in Fig. \ref{fig_ActivePowerControl}, is implemented as a user-defined model (UDM). It is assumed that the parameters in the active power control loop are the same for all WTGs and given as follows
\begin{equation*}
\begin{split}
& T_{\text{sp}}=60,T_{\text{pc}}=0.05,K_{\text{ptrq}}=3,K_{\text{itrq}}=0.6
\end{split}
\end{equation*}
The SSC is implemented by using dynamically linked blocks (DLBs) in C/C++, which allows for advanced control implementation \cite{raoufat2017dynamic}. The overall dynamic simulation is performed in TSAT \cite{DSA_TSAT}.  

The safety limit is set to be 59 Hz. The worse-case contingency is the loss of an entire traditional plant, unit 7, which is a 400 MW generation loss, at one second. Since the SSC aims at preventing unnecessary UFLS, the worst-case contingency with respect to the SSC design is assessed in a way that the steady-state frequency is safe. In addition, the security of other variables like voltages and stability should be guaranteed in the post-fault steady state as well.\par

\begin{figure}[h]
	\centering
	\includegraphics[scale=0.4]{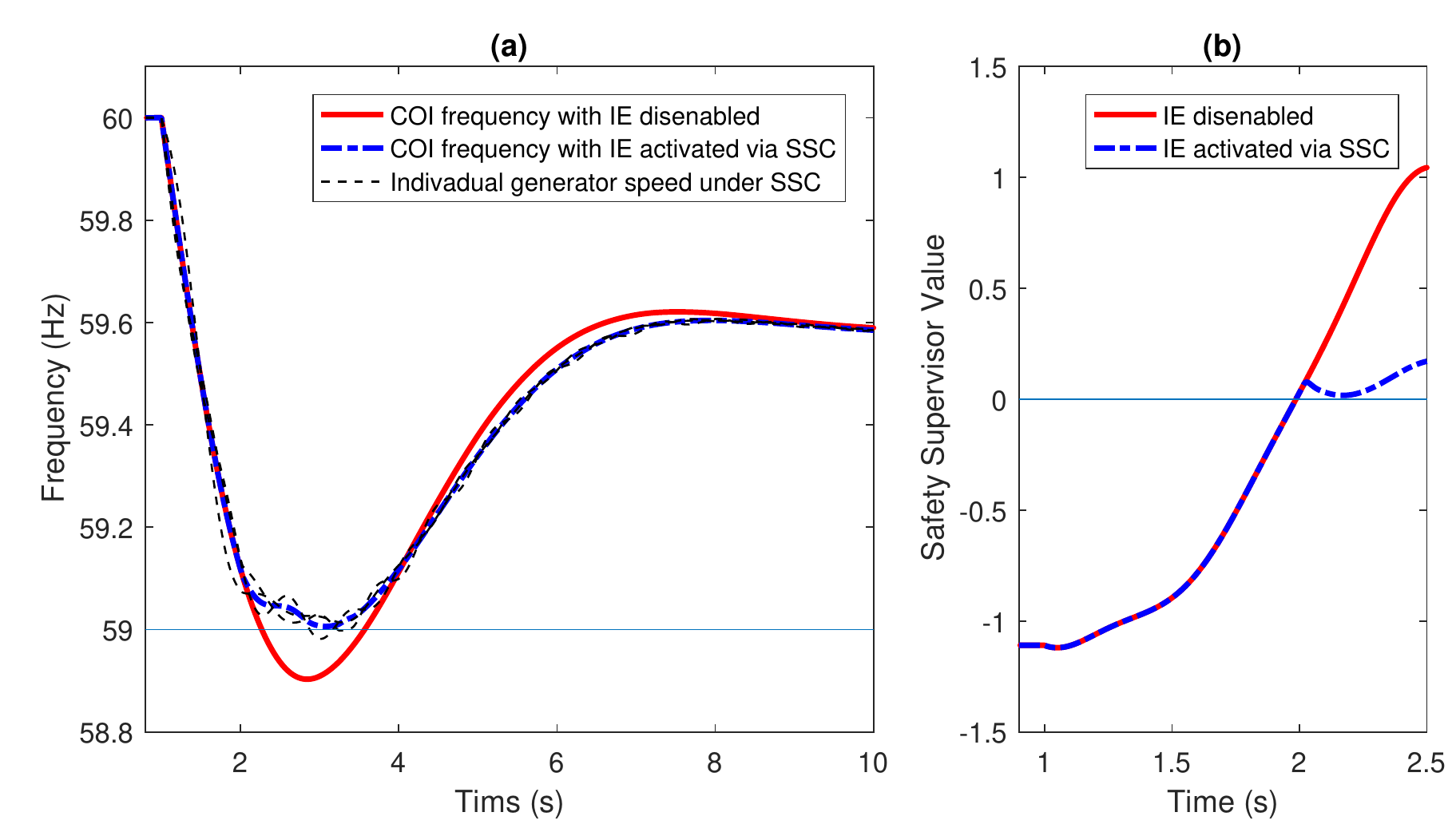}
	\caption{Frequency response under no inertia emulation and inertia emulation activated via safety supervisory control (SSC), (a) frequency response, (b) value of safety supervisor. The reaction time is around one second in this scenario.}
	\label{fig_Performance1}
\end{figure}

In the first scenario, assume the right-hand side of Eq. (\ref{eq_WTG_gain}) is not binding. Then, WTG 5 is selected with $k_{\text{ie}}=0.2$. Under the worst-case contingency, the COI frequencies of the no switching case and supervised switching one are compared in Fig. \ref{fig_Performance1} (a). As seen, the supervisory control timely activates the IE function of WTG 5 based on the supervisory value (shown in Fig. \ref{fig_Performance1} (b)) so that the COI frequency stays within the specified safety limit. Since the supportive gain is large, there is approximately a one second reaction time for WTG 5 to respond. Individual synchronous generator speeds are also plotted. As seen, they are close to the COI frequency. So ensuring safe COI frequency response will greatly reduce the possibility of unnecessary frequency relay actions. \par

\begin{figure}[h]
	\centering
	\includegraphics[scale=0.4]{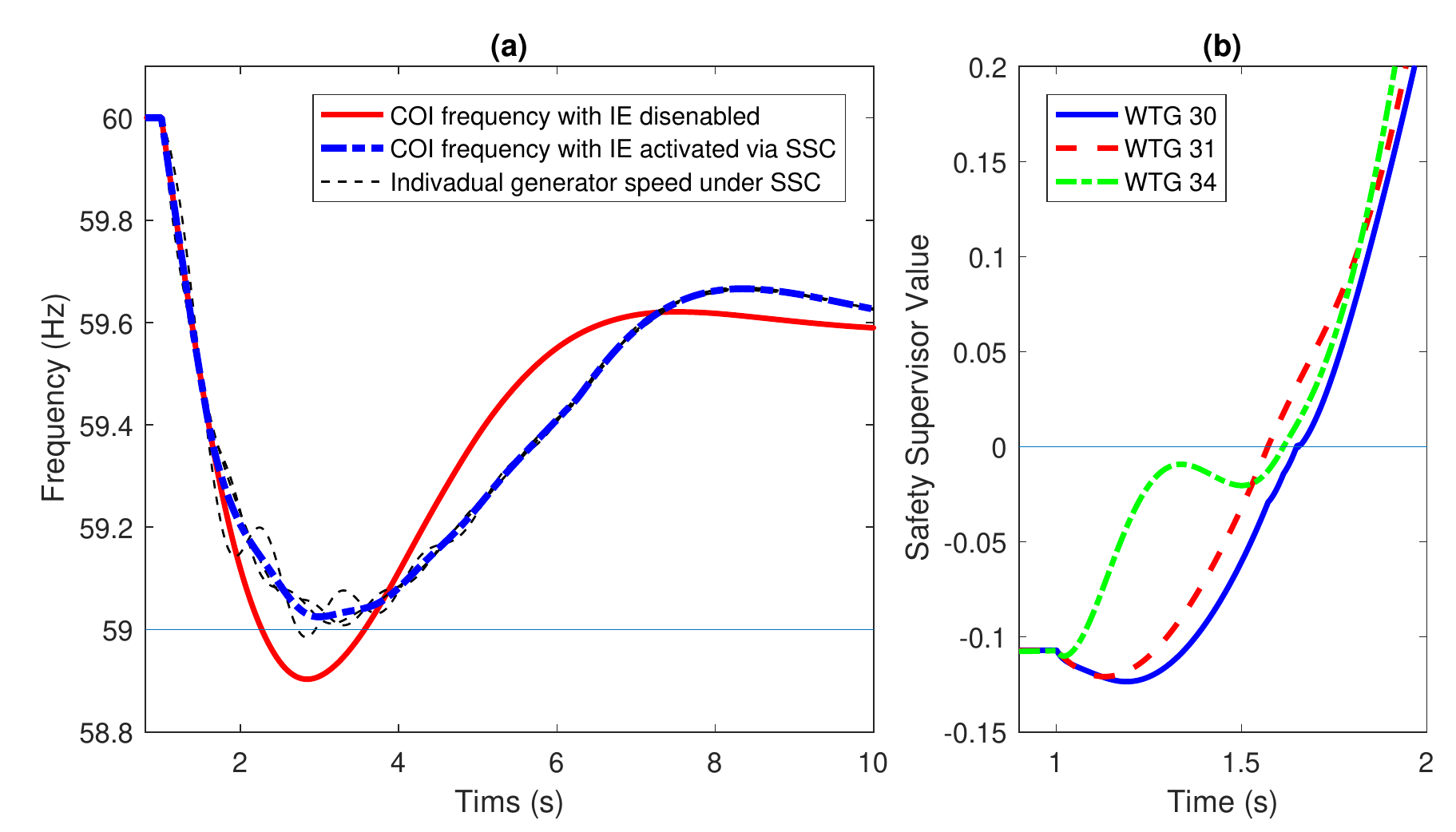}
	\caption{Frequency response under no inertia emulation and inertia emulation activated via safety supervisory control (SSC), (a) frequency response, (b) value of safety supervisor. The reaction time is from 0.5 to 0.7 s in different WTGs in this scenario.}
	\label{fig_Performance3}
\end{figure}

\begin{figure}[!h]
	\centering
	\includegraphics[scale=0.4]{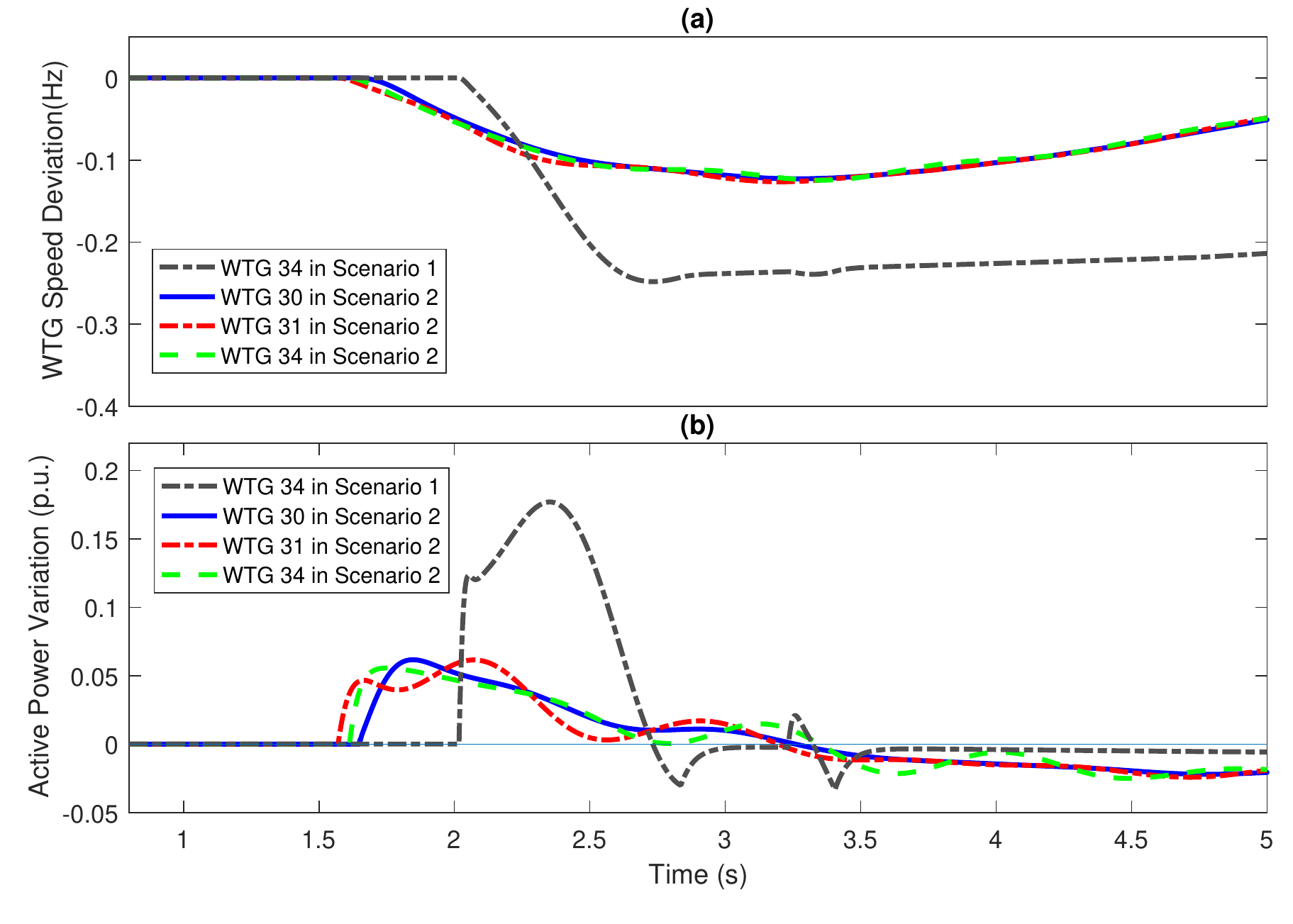}
	\caption{(a) WTG Speeds in different scenarios. (b) Active power variation of WTGs in different scenarios.}
	\label{fig_Performance3_Power}
\end{figure}

In the second scenario, assume the WTG outputs are binding by the right-hand side of Eq. (\ref{eq_WTG_gain}). Then, three WTGs are chosen to be actuators, i.e., $\mathcal{W}=\{1,2,5\}$, with $k_{\text{ie}}=0.03$. The same contingency is applied. The COI frequency and frequencies of individual synchronous generators are plotted in Fig. \ref{fig_Performance3} (a). Fig. \ref{fig_Performance3} (b) indicates that the IE function is activated at slightly different times, from 0.5 to 0.7 s in the different WTGs. This is because the slight difference in the local frequencies. The speeds and power outputs of WTG 1, 2 and 5 are shown in Fig. \ref{fig_Performance3_Power}. Each of them contributes less than 0.1 per unit deviation from their nominal operating points, while WTG 5 contributes more than 0.15 per unit.  \par

\subsection{Adaptive SSC against Varying Renewable Penetration}
\begin{figure}[h]
	\centering
	\includegraphics[scale=0.45]{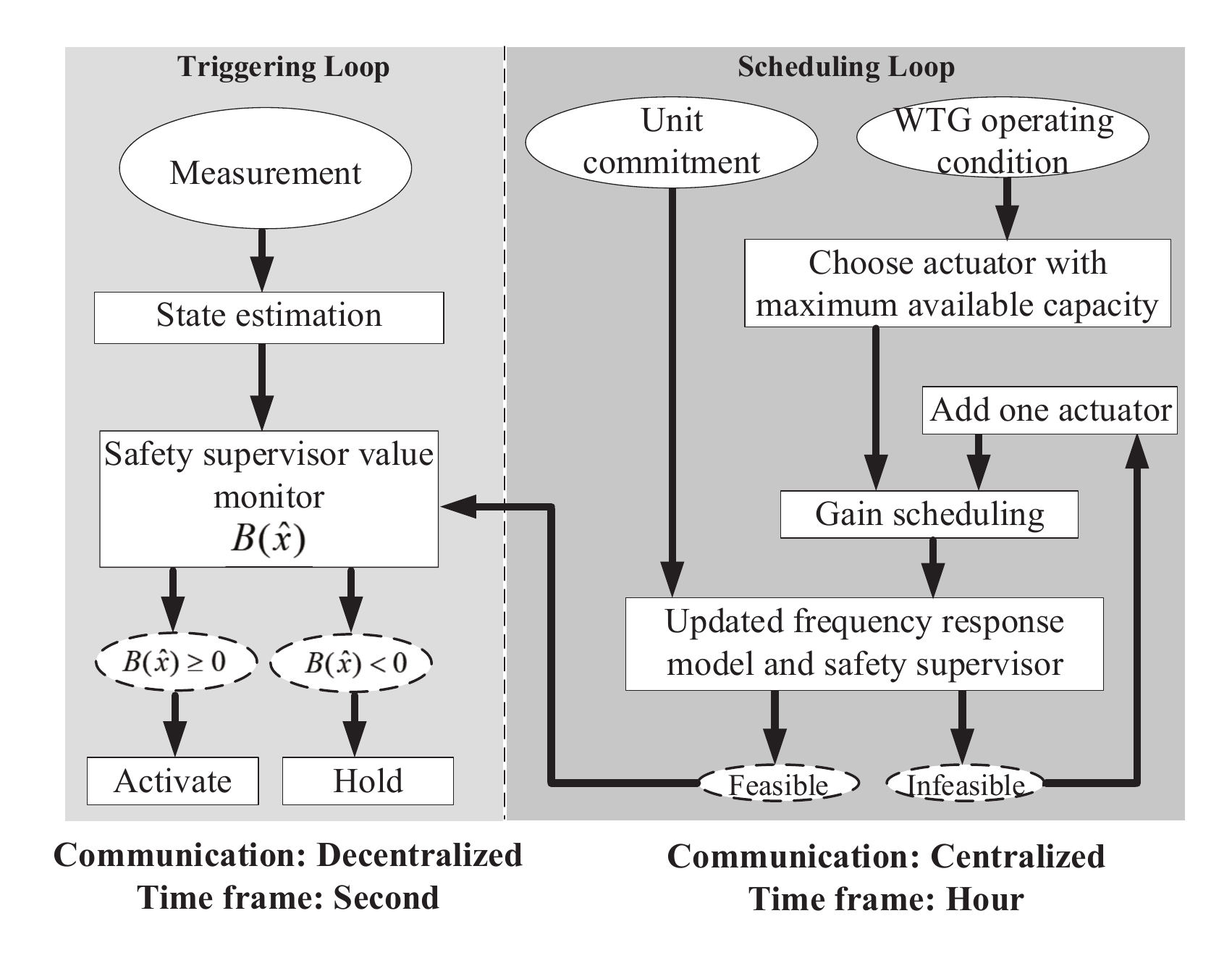}
	\caption{Two-loop SSC with adaptivity and robust to the change of system operating point.}
	\label{fig_TwoLoop}
\end{figure}

Due to the stochastic and intermittent nature of renewable resources, the commitment of traditional plants can change dramatically over time, which could significantly change the system frequency response characteristics. This time-varying feature requires the SSC to be adaptive to the system operating condition. This adaptivity can be implemented by adding a scheduling loop overseeing the triggering loop as shown in Fig. \ref{fig_TwoLoop}. The triggering loop will receive local measurements and make a decision based on the up-to-date supervisor. On the other hand, the scheduling loop will receive global information, such as, unit commitment and WTG outputs, and then recalculate settings for the safety supervisor. When choosing actuators, those with larger available capacity will be selected first. Then, based on the resource availability, the IE gain will be scheduled according to Eq. (\ref{eq_WTG_gain}). The SFR model will be updated and the supervisor will be re-calculated. If the SOS program is not feasible, more WTGs are incorporated and the percentage coefficients in Eq. (\ref{eq_WTG_gain}) will be adjusted.

The scheduling loop will need a centralized communication link. But the two loops are in different time scales. When a disturbance takes place, the SSC uses the latest received ROS as the threshold function to determine the activation of the IE mode. Therefore, the triggering level stays in a decentralized fashion. This is importance since the time scale of this level is in terms of seconds. The scheduling level will be in the same time scale of economic dispatch, and can be regarded as an enhanced functionality of energy management system.

\begin{figure}[h]
	\centering
	\includegraphics[scale=0.4]{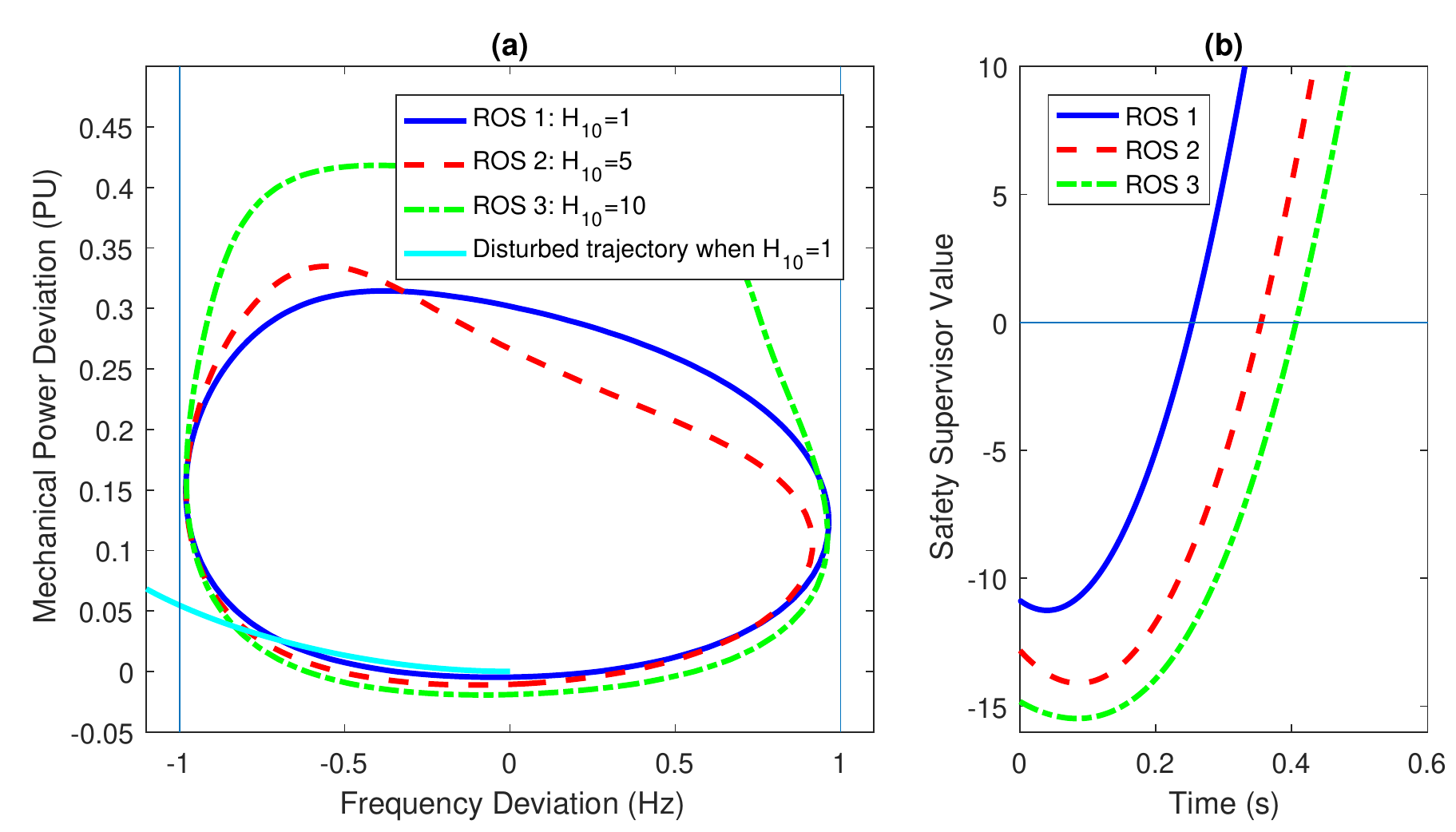}
	\caption{(a) ROSs under different levels of renewables. The ROS shrinks with the increase level of renewable penetration. (b) Values of different supervisors with respect to the disturbed trajectory when $H_{10}=1$ s.}
	\label{fig_AdaptiveSSC}
\end{figure}

\begin{figure}[!h]
	\centering
	\includegraphics[scale=0.4]{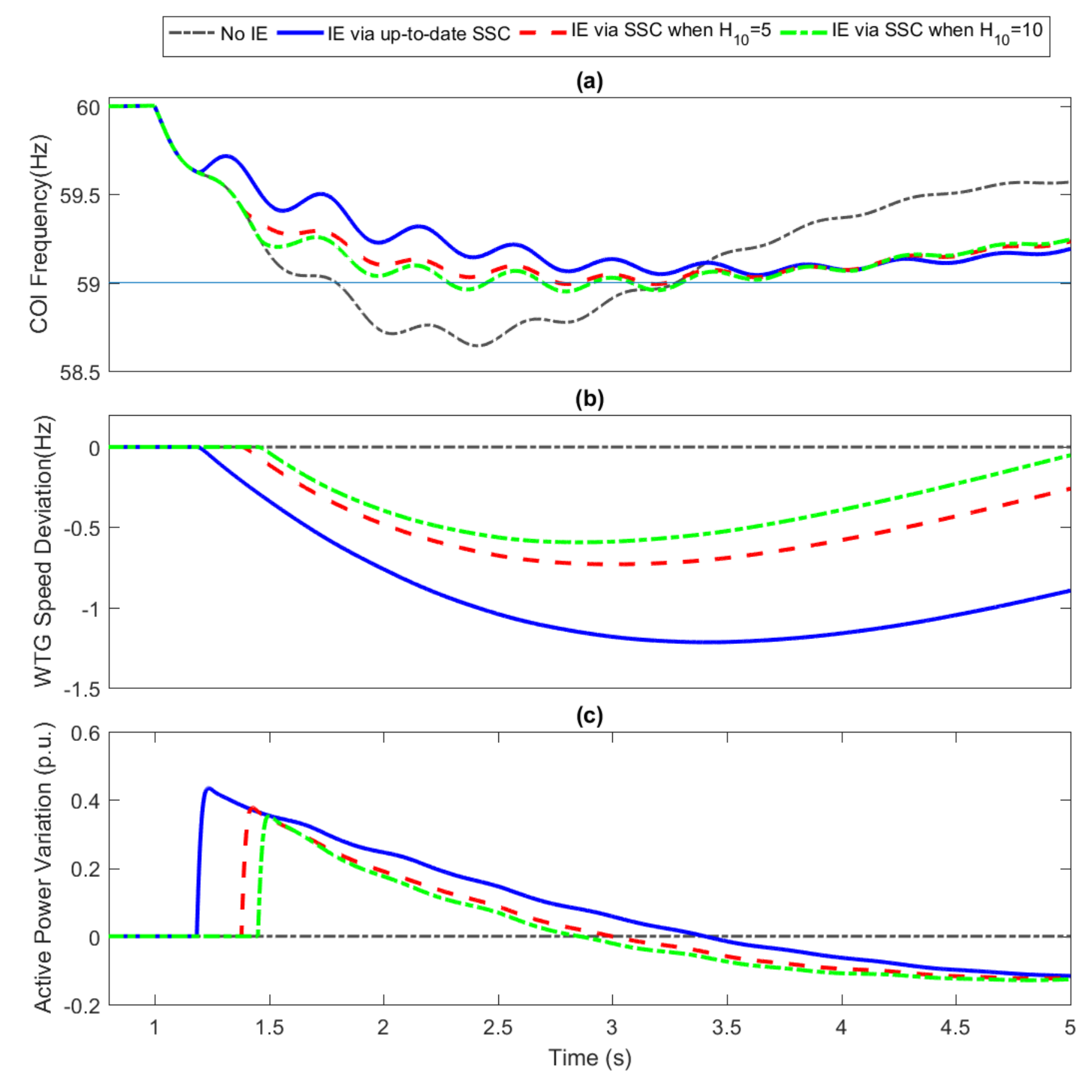}
	\caption{(a) COI Frequencies under different SSCs. (b) WTG Speeds under different SSCs. (c) Active power variation of WTGs under different SSCs.}
	\label{fig_Performance_Ad}
\end{figure}

The demonstration here is based on the setup of Scenario 1, that is, $\mathcal{W}={5}$ and $k_{\text{ie}}=0.2$. The worst-case disturbance and safety limit are also the same. In New England system, SG 10 is to equivalently model the rest of the Eastern Interconnections. Assume a scenario where the level of renewable penetrations in the Eastern Interconnections increases significantly within a time. This change can be equivalently represented by decreasing the inertia constant of SG 10. Here, three different constants, that is, 10, 5 and 1, are used to represent different unit commitment scenarios at certain time snapshots. Based on the information, the scheduling loop will update the SFR model and re-calculate the ROS. Thus, three different ROSs will be obtained with respect to the three inertia constants of SG 10 shown in Fig. \ref{fig_AdaptiveSSC} (a). The ROS shrinks with the increase level of renewable penetration. Now assume the worse-case disturbance happens when $H_{10}=1$. Based on up-to-date ROS 1, which corresponds to the scenario of $H_{10}=1$, the adequate reaction time should be around 0.2 s as shown in Fig. \ref{fig_AdaptiveSSC} (b), and the safety of COI frequency can be ensured shown in Fig. \ref{fig_Performance_Ad} (a). If not updated in time, that is, either ROS 2 or 3 is online, the IE will not be activated in time, and the corresponding COI frequencies are not safe also depicted in Fig. \ref{fig_Performance_Ad} (a). The speeds and outputs of WTGs under up-to-date and out-of-date SSCs are also plotted in \ref{fig_Performance_Ad} (b) and (c), respectively. 

\subsection{Discussion}

As presented, the SSC provides reaction time to critical margin by the level set value of the supervisor. This answer to the question of when to switch is general as the SSC is designed analytically and systematically. The fact that the nadir of the obtained COI frequency response is very close to the safety limit shown in Figs. \ref{fig_ROS1}, \ref{fig_Performance1} and \ref{fig_Performance3} indicates that the estimated LROS is highly precise. In other words, the critical switching instant, or equivalent largest deadband, has been successfully computed. On the one hand, these results verify the proposed formulation. On the other hand, larger deadband filters out frequency fluctuation and ensure the response of WTGs to only sizable disturbances. Thus, the support action induced mechanical stress of WTGs can be minimized.

It is also worth mentioning that although the type-3 WTG is adopted for illustration, the SSC can be applied to any type of converter-interfaced sources, including but not limited to the type-4 WTG, photovoltaic generators, and energy storage systems. When a system is supplied dominantly by wind, the kinetic energy in WTGs is off-the-shelf compared to the energy storage system. Such a configuration is also available for synthesis of switching actions in remedial action schemes \cite{zyc_RAS}.

The RoCoF is another major factor for frequency relay actions. Since the largest RoCoF generally arises at the first moment after a disturbance takes place, it is very difficult or even impossible to improve it in a corrective fashion, which is the focus of this paper. In other words, RoCoF should be addressed in a preventive way. Commitment of traditional synchronous generators or new fast-response devices is the effective approach to limit the RoCoF \cite{Prakash2017}\cite{Xu2018}.

\section{Summary}\label{sec_con}
This paper first interprets the mode synthesis principle for safe response and the concept of ROS. Then, a mathematical optimization problem in the functional space is proposed to estimate the LROS. The optimization problem is explained from a geometric point of view, and then converted into a SOS program by using polynomial functions and semi-algebraic sets. A feasible result of the SOS program will generate a barrier certificate. The superlevel set of the barrier certificate over-approximates the backward reachable set of the unsafe set and the sublevel set of it under-approximates the ROS. This barrier certificate is employed as the safety supervisor for hybrid supportive mode synthesis of WTGs. The proposed controller is verified on a single-machine three-phase nonlinear microgrid model in Simulink. For multi-machine systems, a decentralized SSC is designed for small-scale systems and demonstrated on the IEEE 39-bus system with high renewables modeled in DSATools. Both results indicate that the proposed framework and control configuration will ensure adequate response with relatively little conservativeness. Finally, a scheduling loop is proposed so that the supervisor updates its boundary with respect to the renewable penetration level so as to be robust against variations in system inertia. The shape change of ROSs with respect to renewable penetration level is demonstrated. Future work will focus on reducing computational complexity by using alternative methods rather than SOS decomposition so that higher order analytical models can be employed. In addition, a comprehensive supportive function including inertial emulation, primary response and safety recovery with de-loaded WTGs using the SSC will be studied.

\appendices
\section{Discussion on Gain of Inertia Emulation}\label{apen_gain}
The equation in (\ref{eq_WTG_gain}) is derived as follows. First, the supportive power from WTG $i$ can be approximately regarded in proportion to the averaged value of RoCoF $A_{r}$, that is, $\Delta P_{g,i}=S_{i}^{w}k_{\text{ie}}A_{r}$, which will comply with the WTG operating limit as
\begin{equation}
\label{eq_WTG_limit}
S_{i}^{w}k_{\text{ie}}A_{r}\leq \min\{P_{\text{max}}-p_{g},p_{g}-P_{\text{min}}\} \text{ [W]}
\end{equation}
The first term on the right hand side denotes the WTG output power limit. The second term equivalently represents the rotor speed security limit. The calculation of $P_{\text{min}}$ can be found in \cite{Wang2018Novel}.

As well known, the actual inertial response can be calculated as follow
\begin{equation}
\Delta p_{\text{ir}}(t)=\Delta P_{d}-\Delta p_{m}(t) \text{  [W]}
\end{equation}
Up to the instant of frequency nadir, $p_{m}(t)$ can be approximated using a linear relation as $\Delta p_{m}(t)=A_{m}t$ as shown in \cite{Baldick2014}. Averaging the right-hand side of the above equation during $T_{\text{nad}}$ yields the averaged inertial response as follows
\begin{equation}
\Delta P_{\text{IR}}=\Delta P_{d}-0.5A_{m}T_{\text{nad}} \text{  [W]}
\end{equation}
Assume that the frequency deviation can be limited if $k_{\text{ad}}$ percentage of $\Delta P_{\text{IR}}$ is compensated by WTGs, and WTG $i$ can contribute to a $\rho_{i}$ percentage of the total requirement. Then, we will have
\begin{equation}
\label{eq_WTG_output}
\Delta P_{g,i}=k_{\text{ie}}A_{r}=\rho_{i}k_{\text{ad}}\Delta P_{\text{IR}}=\rho(\Delta P_{d}-0.5A_{m}T_{\text{nad}}) \text{  [W]}
\end{equation}
Combining Eq. (\ref{eq_WTG_limit}) and (\ref{eq_WTG_output}) yields Eq. (\ref{eq_WTG_gain}), where $k_{\text{ad}}$, $A_{r}$ and $A_{m}$ will be adjusted via trial-and-error procedures, and $\rho_{i}$ can be determined by a scheduling algorithm once global information of all WTG outputs is received such that the summation of $\rho_{i}$ will be $100\%$. In conclusion, once the worst-case contingency $\Delta P_{d}$ is given, the controller gains are a function of number of actuators and their current outputs.

\bibliography{IEEEabrv_zyc,library,Ref_SSC}  
\bibliographystyle{IEEEtran}

\end{document}